\documentclass[amsmath,amssymb, aps,reprint, groupedaddress, onecolumn,notitlepage, superscriptaddress,longbibliography,nofootinbib,showpacs]{revtex4-1}
\usepackage{amsmath}
\usepackage{amssymb}
\usepackage{multirow}
\usepackage{verbatim}
\usepackage{color,graphicx}

\usepackage{amsfonts}   % if you want the fonts
\usepackage{setspace}
\usepackage{dsfont}
\usepackage[all]{xy}
\usepackage[margin=1.5in]{geometry}  % set the margins to 1in on all sides
\usepackage{amssymb}
\usepackage{graphicx}              % to include figures
\usepackage{amsmath}               % great math stuff
\usepackage{amsfonts}              % for blackboard bold, etc
\usepackage{amsthm}                % better theorem environments
\usepackage{algpseudocode}  %algorithm
\usepackage{algorithm}
\usepackage{graphicx}
\usepackage{caption}
\usepackage{subcaption}
\usepackage{rotating}
\usepackage{color}
\usepackage[toc,page]{appendix}
\usepackage[english]{babel}
\usepackage{lineno, blindtext}
\usepackage{appendix}
\usepackage[nottoc,numbib]{tocbibind}
\usepackage{epstopdf}
\usepackage{bbold}

\usepackage[colorlinks]{hyperref}

\theoremstyle{definition}

\newtheorem{rmk}{Remark}[section]

\newcommand{\RR}{\mathbb{R}}      % for Real numbers
\newcommand{\CC}{\mathbb{C}}      % for Real numbers

\newcommand{\vecc}{\boldsymbol}

\begin{document}

\title{\bf On the Small Mass Limit of Quantum Brownian Motion with Inhomogeneous Damping and Diffusion}


\author{Soon Hoe Lim} \email[]{shoelim@math.arizona.edu} \affiliation{Department of Mathematics, University of Arizona, Tucson, AZ 85721, USA} \affiliation{Program in Applied Mathematics, University of Arizona, Tucson, AZ 85721, USA}
\author{Jan Wehr} \affiliation{Department of Mathematics, University of Arizona, Tucson, AZ 85721, USA} \affiliation{Program in Applied Mathematics, University of Arizona, Tucson, AZ 85721, USA} \author{Aniello Lampo} \affiliation{ICFO - Institut de Ciencies Fotoniques, The Barcelona Institute of Science and Technology, 08860 Castelldefels (Barcelona), Spain}  \author{Miguel \'Angel Garc\'{\i}a-March} \affiliation{ICFO - Institut de Ciencies Fotoniques, The Barcelona Institute of Science and Technology, 08860 Castelldefels (Barcelona), Spain}  \author{Maciej Lewenstein} \affiliation{ICFO - Institut de Ciencies Fotoniques, The Barcelona Institute of Science and Technology, 08860 Castelldefels (Barcelona), Spain} \affiliation{ICREA - Instituci{\'o} Catalana de Recerca i Estudis Avan{\c c}ats, Lluis Companys 23, E-08010 Barcelona, Spain}

\date{\today}

\begin{abstract}
We study the small mass limit  (or: the Smoluchowski-Kramers limit) of a class of quantum Brownian motions  with inhomogeneous damping and diffusion. For Ohmic bath spectral density with a Lorentz-Drude cutoff, we derive the Heisenberg-Langevin equations for the particle's observables using a quantum stochastic calculus approach. We set the mass of the particle to equal $m = m_{0} \epsilon$, the reduced Planck constant to equal $\hbar = \epsilon$ and the cutoff frequency to equal $\Lambda = E_{\Lambda}/\epsilon$, where $m_0$ and $E_{\Lambda}$ are positive constants, so that the particle's de Broglie wavelength and the largest energy scale of the bath are fixed as $\epsilon \to 0$. We study the limit as $\epsilon \to 0$ of the rescaled model and derive a limiting equation for the (slow) particle's position variable. We find that the limiting equation contains several drift correction terms, the {\it quantum noise-induced drifts}, including terms of purely quantum nature, with no classical counterparts.
 %\PACS{PACS code1 \and PACS code2 \and more}
\end{abstract}

\maketitle

\section{Introduction}
Multiscale analysis of both classical and quantum systems has been a subject of active investigation in recent years. The underlying idea is that due to the presence of widely separated characteristic time scales in the system, one can obtain a simplified, reduced model that often captures the essential dynamics on a coarse-grained time scale \cite{bo2016multiple, Pavliotis, givon2004extracting, volpe2016effective,berglund2006noise}. Depending on the nature of the systems, different approaches can be undertaken to implement this idea. For instance, Markovian limits such as weak coupling limit and repeated interaction limit were studied in \cite{ accardi2002quantum,Dereziński2008, bouten2015trotter,gough2005quantum,dhahrimarkovian,de2010quantum,de2014quantum}
to justify the use of effective equations such as quantum Langevin equations in modeling quantum systems arising in quantum optics and quantum electrodynamics \cite{cohen1992atom,haroche2006exploring,GardinerBook}. Adiabatic elimination type problems for open quantum systems were studied in \cite{Haake-ZfP,haake1983adiabatic,bouten2008adiabatic,bouten2008approximation,gough2007singular,petersen2010singular,vcernotik2015adiabatic,azouit2016adiabatic,gough2014zeno} and perturbative methods were considered in \cite{reiter2012effective,kessler2012generalized,li2014perturbative,rivas2017refined}.

Of particular interest is the small mass limit (or the Smoluchoswki-Kramers limit \cite{Smoluchowski, Kramers}) of noisy dynamical systems. It has been extensively studied and is well understood for classical systems; see for instance, \cite{Hottovy2012,hottovy2015smoluchowski,herzog2016small,birrell2017small,2016BirrellarXiv160808194B,LimWehr_Homog_NonMarkovian,birrell2017}. On the other hand, analogous questions for quantum systems \cite{Weiss, caldeira2014introduction} are more intricate, and there were few attempts to study the small mass limit, or the related strong friction limit for quantum systems. Such study was initiated and refined in the series of works \cite{pechukas2000quantum, pechukas2001quantum, Ankerhold1,ankerhold2003phase,ankerhold2005quantum} for the Caldeira-Leggett model of quantum Brownian motion (QBM) \cite{Caldeira1983a,SchlosshauerBook,Massignan2015,ferialdi2017dissipation}. In these works, a quantum Smoluchowski equation, an equation for the coordinate-diagonal elements of the density operator (i.e. the position probability distribution), was derived in the overdamped regime. 

The results of these works (see for instance, \cite{pechukas2000quantum}) say that  the strong friction limit of quantum mechanics is essentially classical mechanics as the quantum effects are buried in the fast momentum variable, which is adiabatically eliminated due to separation of associated time scales in the limit. Such limit is the opposite of the weak coupling limit \cite{BreuerBook,Rivas2012}, and its result can be viewed as a consequence of decoherence due to the strong coupling. In \cite{coffey2006semiclassical, luczka2005diffusion,  Maier2010,2017arXiv170104084J}, more careful analysis and related applications were presented, whereas in   \cite{dillenschneider2009quantum} a Heisenberg approach was used. All these attempts rely on asymptotic expansions to study a restricted class of QBM, where the coupling of the system to the environment is linear in the system's position. One important message from these asymptotic expansions is that  the leading correction term to the Smoluchowski  equation is a quantum correction that dominates the classical ones in the low temperature regime, revealing the important role of zero-point quantum fluctuations. 

Similar studies for QBM in inhomogeneous environments are even more interesting \cite{buttiker1987transport}. Such study was conducted in \cite{Barik05,bhattacharya2011phase}, where a semi-classical Langevin approach was employed. In \cite{Sancho1982}, using the Fokker-Planck equation, the authors derived a limiting $c$-number Langevin equation for the position variable in the overdamped limit. While the limiting equation obtained contains interesting quantum correction terms, these studies are not satisfactory for two main reasons. First, an ad-hoc Markovian approximation is made before the overdamped limit was studied. Second, a semi-classical approach is used and assumed that the quantum fluctuations around the mean value of the system's observables are small. Therefore, a more systematic study that takes into account the full quantum nature of the model, including the noise, is necessary. Motivated by this and our goal to generalize the study of small mass limit to quantum dissipative systems, this paper presents a {\it quantum stochastic calculus} approach to study a related limit for a class of non-Markovian QBM with inhomogeneous damping and diffusion. In particular, we will model the noise using the fundamental noise processes of the theory of quantum stochastic calculus, introduced by Hudson and Parthasarathy in the seminal paper \cite{hudson1984quantum}. Rigorous justification of the results in this paper will be presented elsewhere. 

The paper is organized as follows. In Section \ref{model}, we introduce a QBM field model to model interaction of a quantum particle with an equilibrium quantum heat bath at a positive temperature. In Section \ref{approach},  we present the exact Heisenberg equations of motion for the particle's observables. In Section \ref{background_qsde}, we review some basic results from Hudson-Parthasarathy (H-P) quantum stochastic calculus and fix our notations. Modeling the stochastic force, appearing in the Heisenberg equations, using Hudson-Parthasarathy quantum noise processes,  we derive a quantum stochastic differential equation (QSDE) version of the QBM model in Section \ref{qnoise}. We identify the characteristic time scales of the model and study its rescaled version in Section \ref{rescaledmodel}. Our main result is the derivation of the effective equation (see eqn. \eqref{mainresult}) for the (slow) position variable in the limit as all the characteristic time scales of the model tend to zero at the same  rate. The derivations, as well as discussions of the results, are presented in Section \ref{derive}. We end the paper by stating the conclusions and making some remarks in Section \ref{final}. 

%-------------------------------------------

\section{QBM Model} \label{model}
In this section, we introduce a one-dimensional Hamiltonian model to study the dynamics of a quantum Brownian particle coupled to an equilibrium heat bath. The particle interacts with the heat bath via a coupling, which is a function of the position variables. This function can be nonlinear in the system's position, in which case the particle is subject to inhomogeneous damping and diffusion \cite{Weiss,Massignan2015,lampo2016lindblad}.  The model can be viewed as a field version of the one studied in \cite{Massignan2015,lampo2016lindblad}, a generalization of the Pauli-Fierz model \cite{pauli1938theorie, derezinski1999asymptotic,derezinski2001spectral}, or a quantum analog of the one studied in Appendix A of \cite{LimWehr_Homog_NonMarkovian}. It is a fundamental model which not only allows simple analytic treatments and provides physical insights, but also realistically models many open qantum systems --- for instance, an atom in an electromagnetic field. 

%We describe the particle, the heat bath and their interaction in the model more precisely in the following. The Brownian particle is a quantum mechanical system, denoted $\mathcal{S}$, with energy operator $H_{S}$ on the  Hilbert space $\mathcal{H}_{S} := L^2(\RR)$. It  is subjected to a confining, smooth potential $U(X)$.  The infinite heat bath, denoted $\mathcal{B}$, is a field of mass-less bosons at a positive temperature. It is described by the triple $(\mathcal{H}_{B}, \rho_{\beta}, H_{B} )$, where $\mathcal{H}_{B} := \Gamma(L^2(\RR^+))$, $\RR^+ = [0,\infty)$, is the bosonic Fock space over $L^2(\RR^+)$, $H_{B}$ is the Hamiltonian of the heat bath defined on $\mathcal{H}_{B}$ and $\rho_{\beta} = e^{-\beta H_B}/Tr(e^{-\beta H_B})$ is the Gibbs thermal state at an inverse temperature $\beta = 1/(k_{B}T)$. We take $H_{B} = d\Gamma(H_B^1)$, the differential second quantization of the energy operator $H_{B}^{1}$ which acts in the one-particle frequency space $L^2(\RR^+)$ as: \begin{equation}(H_{B}^{1} \phi)(w) = \epsilon(w) \phi(w),\begin{equation} where  $\epsilon(w)$ is the energy of a boson with frequency $w \in \RR^{+}$. The function $\epsilon(w)$ is the dispersion relation for the bath, which in our case, is a linear one, i.e. $\epsilon(w) = \hbar w$.  The equilibrium frequency distribution of bosons at an inverse temperature $\beta$ is given by the Planck's law:  \begin{equation}\nu(w) = \frac{1}{\exp{\left( \beta \epsilon(w) \right)}-1}.\begin{equation}

The full dynamics of the model is described by the Hamiltonian: 
%on the total Hilbert space $\mathcal{H}_{S} \otimes \mathcal{H}_{B}$: 
\begin{equation}
H = H_{S} \otimes I + I \otimes H_{B} + H_{I} + H_{ren} \otimes I, \end{equation} where $H_S$ and $H_B$ are Hamiltonians for the particle and the heat bath respectively, given by
\begin{equation}
H_{S} = \frac{P^2}{2m} + U(X), \ \ \ H_{B} = \int_{\RR^{+}} \hbar \omega   b^{\dagger}(\omega)b(\omega) d\omega,\end{equation}
$H_{I}$ is the interaction Hamiltonian given by
\begin{equation}
H_{I} = - f(X) \otimes \int_{\RR^{+}} [c(\omega) b^{\dagger}(\omega)+\overline{c(\omega)}b(\omega) ] d\omega, \end{equation} 
and $H_{ren}$ is the renormalization Hamiltonian given by 
\begin{equation}
H_{ren} = \left( \int_{\RR^{+}} \frac{|c(\omega)|^2}{\hbar \omega} d\omega \right) f(X)^2. \end{equation}
Here $X$ and $P$ are the particle's position and momentum operators,  $m$ is the mass of the particle, $U(X)$ is a smooth confining potential, $b(\omega) $ and $b^{\dagger}(\omega)$ are the bosonic annihilation and creation operator of the boson of frequency $\omega$ respectively and they satisfy the usual canonical commutation relations (CCR): $[b(\omega),b^{\dagger}(\omega')] = \delta(\omega-\omega'), \ \ [b(\omega),b(\omega')]=[b^{\dagger}(\omega), b^{\dagger}(\omega')] = 0.$  We assume that the operator-valued function $f(X)$ is positive and can be expanded in a power series, and $c(\omega)$ is a complex-valued coupling function (form factor) that specifies the strength of the interaction with each frequency of the bath.  It determines the spectral density of the bath and therefore the model for damping and diffusion of the particle. The heat bath is initially in the Gibbs thermal state, $\rho_{\beta} = e^{-\beta H_B}/Tr(e^{-\beta H_B})$,  at an inverse temperature $\beta = 1/(k_{B}T)$.

In this paper, we consider the coupling function: 
%two coupling function with different cutoffs - Ohmic at low freqs, but one corresponds to markovian noise and other not markovian (so cutoff choice is important) 
%nid are different in both cases, in contrast to classical case when fdt holds - reason: in quantum case the drift induced by the zero point fluctuations depend in detail on the cutoff (even if qfdt holds!) the contributions from vanishing memory time still cancel off and the contribution from vanishing inertia time also depends on cutoff 
%--- must reduce to the earlier nid and classical nid in certain limits 
%comment on non-Ohmic case, in general cannot be sum of lorenztian (difference of these will get us difference of two deltas that cancel in the limit) 
%approximate power law by sum of exps  
\begin{equation} \label{coupling}
c(\omega) = \sqrt{\frac{\hbar \omega }{\pi}\frac{\Lambda^2 }{\omega^2+\Lambda^2}},
\end{equation}
where  $\Lambda$ is a positive constant.  The bath spectral density is  given by:
\begin{equation}
J(\omega) := \frac{|c(\omega)|^2}{\hbar} = \frac{\omega}{\pi}\frac{\Lambda^2 }{\omega^2+\Lambda^2}, \end{equation}
which is the well-known Ohmic spectral density with a Lorentz-Drude cutoff \cite{BreuerBook}. The renormalization potential $H_{ren}$ is needed to ensure that the bare potential acting on the particle is $U(X)$ and that the Hamiltonian can be written in a positively defined form: $H=H_{S} \otimes I + H_{B-I}$, where $H_{B-I}$ is given by
\begin{equation} \label{trans_inv}
H_{B-I} = \int_{\RR^{+}} \hbar \omega \left(b(\omega)-\frac{c(\omega)}{\hbar \omega} f(X) \right)^{\dagger} \left(b(\omega)-\frac{c(\omega)}{\hbar \omega} f(X) \right) d\omega. \end{equation}

%----------------------------------------
\section{Heisenberg-Langevin Equations} \label{approach}

In this section, we present the Heisenberg equations of motion and  study  the stochastic force term appearing in the equation. This will pave the way to model the action of the heat bath on the particle by appropriate quantum colored noises introduced in the next sections. Our final goal is the construction of dissipative non-Markovian Heisenberg-Langevin equations driven by appropriate thermal noises, which are built from H-P fundamental noise processes. From now on, $I$ denotes identity operator on an understood space and $1_{A}$ denotes  indicator function of the set $A$. 

Define the particle's velocity, $V(t) = \frac{P(t)}{m}$ and note that $f'(X) = -i[f(X),P]/\hbar$.  Starting from the Heisenberg equations of motion and eliminating the bath variables, one obtains the following equations for the particle's observables (see Appendix \ref{appendix_qle} for details of derivations): 
\begin{align}
\dot{X}(t) &= V(t), \label{qle1} \\ 
m  \dot{V}(t) &= -U'(X(t)) - f'(X(t)) \int_{0}^{t} \kappa(t-s) \frac{\{ f'(X(s)), V(s)\}}{2} ds \nonumber \\ 
&\ \ \ \ \ \ + f'(X(t)) \cdot (\zeta(t)-f(X)\kappa(t)), \label{qle2}
\end{align}
where 
\begin{align}
\kappa(t) &= \int_{\RR^{+}} d\omega \frac{2 |c(\omega)|^2}{\hbar \omega} \cos(\omega t)  = \int_{\RR^{+}} d\omega \frac{2 J(\omega)}{\omega} \cos(\omega t) 
\end{align}
is the {\it memory kernel},
\begin{align}
\zeta(t) &= \int_{\RR^{+}} d\omega c(\omega) (b^{\dagger}(\omega)e^{i\omega t}+b(\omega)e^{-i\omega t})   
\end{align} 
is a {\it stochastic force} whose correlation function depends on the coupling function, $c(\omega)$, and the distribution of the initial bath variables, $b(\omega)$ and $b^{\dagger}(\omega)$ -- let us remind that we initially consider a thermal Gibbs state. The term $f'(X(t))f(X) \kappa(t)$ is the initial slip term \cite{MLFH}. The initial position and velocity are given by $X$ and $V$ respectively.

The above equations are exact, non-Markovian, operator-valued and describe completely positive dynamics.  Note that in the damping term which is nonlocal in time, we have an anti-commutator, which does not appear in the corresponding classical equation or in the equation for the linear QBM model (where $f(X)=X$). The presence of the anti-commutator is thus a quantum feature of the inhomogeneous damping.  

The initial preparation of the total system, which fixes the statistical properties of the bath operators and of the system's degrees of freeedom, turns the force $\zeta(t)$ into a random one \cite{hughes2006dynamics}. We  specify a preparation procedure to fix the properties of the stochastic force. To this end, we absorb the initial slip term into the stochastic force, defining:
\begin{equation} \xi(t) := \zeta(t) - f(X) \kappa(t). \end{equation}
With this, in the nonlinear coupling case, the equation for the particle's velocity is driven by the multiplicative noise $f'(X(t)) \xi(t)$. From now on, we refer to $\xi(t)$ as the {\it quantum noise}. The statistics of $\xi(t)$ depends on the distributions of the initial bath variables $(b(\omega), b^{\dagger}(\omega))$ and the initial system variable $f(X)$. 

%The initial slip term  appears also in the absence of the potential renormalization in the Hamiltonian. With this initial value contribution being absorbed into the quantum fluctuation term, these become stationary fluctuations with respect to the initial density operator of the bath  given by $\rho = e^{-\beta (H_{B}+H_{I}+H_{ren})}/Z$ (which should fix the initial preparation of the bath).  Note, however, that with respect to an average over the bare, non-shifted bath density operator, the quantum fluctuations would become non-stationary. 

Denoting by $E_{\beta}$ the expectation with respect to the thermal Gibbs state $\rho_{\beta}$ at the temperature $T$,  we have
\begin{align}
&E_\beta[(b^{\dagger}(\omega)e^{i\omega t}+b(\omega)e^{-i\omega t})  (b^{\dagger}(\omega')e^{i\omega' s}+b(\omega')e^{-i\omega' s})] \nonumber \\ &=\left[ (1+\nu_{\beta}(\omega)) e^{-i\omega(t-s)} + \nu_{\beta}(\omega) e^{i\omega(t-s)} \right] \delta(\omega-\omega'),
\end{align}
where $\nu_{\beta}(\omega)$ is given by the Planck's law \begin{equation} \nu_{\beta}(\omega) = \frac{1}{\exp{\left( \beta \hbar \omega \right)}-1}. \end{equation}

Since we absorbed the initial slip term into the stochastic force, $\xi(t)$ no longer has a stationary correlation when averaged with respect to $\rho_{\beta}$ \cite{hanggi1997generalized}. However, $\xi(t)$ is stationary and Gaussian when conditionally averaged with respect to $\rho'_{\beta} = e^{-\beta H_{B-I}}/Tr(H_{B-I}),$ where $H_{B-I}$  is the quadratic Hamiltonian defined in \eqref{trans_inv} and the average is conditioned on the initial position variable $X$.

%In this case, denoting $E'_{\beta}$ as the conditional expectation with respect to $\rho_{\beta}'$, the symmetric noise correlation function is given by: \begin{equation} \label{araki} E'_\beta\left[ \{\xi(t), \xi(s)\} \right] =  E_\infty\left[ \{ W^{th}_{t}(c), W^{th}_{s}(c)\} \right].\end{equation} In \eqref{araki}, $E_{\infty}$ denotes expectation with respect to a Fock vacuum state (zero-temperature state), $W^{th}_{t}(c) := \int_{\RR^+} dw [ (a^{th,l})^{\dagger}_{t}(c(w)) + a^{th,l}_{t}(c(w)) ] \in \Gamma(L^2(\RR^{+})  \oplus \overline{L^2(\RR^{+})}) $, where \begin{equation}a^{th,l}_t(c) = a_t(\sqrt{\nu+1}c) \otimes I     +   I \otimes a_t^{\dagger}(\sqrt{\nu}\overline{c}).\begin{equation} Here, $(a^{th,l})^{\dagger}_t(c) + a^{th,l}_t(c) $ is the thermal noise in (left) {\it Araki-Woods representation} defined on the double Fock space \cite{araki1963representations,derezinski2006introduction}. Therefore, one sees that the symmetric correlation function of $\xi(t)$ with respect to $\rho_{\beta}'$ equals the symmetric correlation function of $W_{t}^{th}(c)$ with respect to the Fock vacuum state on the double Fock space. Inspired by this idea, 

We will work in a {\it Fock vacuum representation} using the H-P quantum stochastic calculus approach (see Section \ref{background_qsde}). In particular, our goal is to describe the quantum noise as a  quantum stochastic process satisfying certain QSDE such that the symmetric correlation function of the stochastic process with respect to the vacuum state on an enlarged Fock space coincides with that of  $\xi(t)$ with respect to $\rho_{\beta}'$. As a preparation to achieve this goal, we study $E_{\beta}'[\xi(t)\xi(s)]$ in the following. 
We write:
\begin{align} E'_\beta[\xi(t) \xi(s)] &= \int_{\RR^{+}} d\omega \hbar J(\omega) \left(\coth\left(\frac{\hbar \omega}{2k_{B}T}\right)\cos(\omega(t-s)) - i\sin(\omega(t-s)) \right) \\ 
&=: D_{1}(t-s) - i D(t-s), \end{align}
where $D_{1}$ is the noise kernel given by
\begin{equation} D_{1}(t-s) := E'_\beta[\{\xi(t), \xi(s)\}/2],\end{equation} i.e. the symmetric correlation function of $\xi(t)$ with respect to $\rho_{\beta}'$, and $D$ is the dissipation kernel given by \begin{equation} D(t-s) := i E'_\beta[[\xi(t),\xi(s)]/2],\end{equation} which is related to linear susceptibility. Expanding, one gets for small $\hbar$ (or similarly, for large $T$), $E'_\beta[\xi(t) \xi(s)] =  k_{B}T \kappa(t) + O(\hbar),$ which is the classical Einstein's relation.

For our choice of $c(\omega)$ (see \eqref{coupling}), the memory kernel, $\kappa(t)$, is exponentially decaying with decay rate $\Lambda$, i.e. $\kappa(t) = \Lambda e^{-\Lambda t}.$ Moreover, one can compute, for $t>0$: 
\begin{align} \label{D1}
D_{1}(t) 
&= \frac{\hbar \Lambda}{2} \cot \left(\frac{\hbar \Lambda}{2k_{B}T} \right) \kappa(t) +  \sum_{n=1}^{\infty} \frac{2k_{ B}T \Lambda^2 \nu_{n}}{ \nu_{n}^2 - \Lambda^2} e^{-\nu_{n}t}, 
\end{align}
where $\nu_{n} = \frac{2 \pi n k_{B}T}{\hbar}$ are the bath Matsubara frequencies \cite{carlesso2016quantum}. Also, the dissipation kernel is 
\begin{equation} \label{D}
D(t) = \frac{\hbar \Lambda^3}{2} e^{-\Lambda t}. 
\end{equation}
In this paper, we consider the case $k_B T > \hbar \Lambda/\pi$, so that $\cot(\hbar \Lambda/2k_BT)$ and all the terms in the series in \eqref{D1} are positive. 

We end this section with the following remark. For $T \to 0$ we have instead: 
\begin{equation} E'_{\beta}[\{\xi(t), \xi(s)\}/2] \to  \frac{-\hbar \Lambda^2}{2 \pi} (e^{-\Lambda (t-s)} \overline{Ei}(\Lambda (t-s)) + e^{\Lambda(t-s)} Ei(-\Lambda (t-s))), \end{equation}
where $Ei$ is the exponential integral function defined as follows: 
\begin{equation} -Ei(-x) = \hat{\gamma}(0,x) = \int_{x}^{\infty} e^{-t}/t dt. \end{equation} Here $\overline{Ei}(x) = \frac{1}{2}(Ei^{+}(x) + Ei^{-}(x))$, $Ei^{+}(x) = Ei(x+i0)$, $Ei^{-}(x) = Ei(x-i0)$. The symmetric correlation function obtained above can be interpreted as follows. As the temperature $T$ decreases, the Matsubara frequencies $\nu_{n}$ get closer to each other, so at zero temperature all of them contribute and the sum in \eqref{D1} may be replaced by an integral, which turns out to have expression in terms of the $Ei$ functions \cite{ingold}. In fact, in this case the symmetric correlation function decays polynomially for large times  \cite{jung1985long}. In other words, systems at zero temperature are strongly non-Markovian and the techniques in this  paper cannot be applied to study them.

%------------------------------------------

\section{Preliminaries on Quantum Stochastic Calculus}  \label{background_qsde}
To ensure that our exposition is self-contained, as well as to fix the notations and terminologies, we review some basic ideas and results from H-P quantum stochastic calculus, which is a  boson Fock space stochastic calculus based on the creation, conservation and annihilation operators of free field theory. For details of the calculus, we refer to the monographs \cite{meyer2006quantum} and \cite{parthasarathy2012introduction}. For recent developments, perspectives and applications of the calculus to the study of open quantum systems, we refer to \cite{Hudson1985,fagnola1999quantum,attal2006open, bouten2007introduction, biane2010ito,2017arXiv170500520P,nurdinlinear,emzir2016physical,barchielli2015quantum,gregoratti2001hamiltonian,gough2009quantum,sinha2007quantum}. In the following, we use Dirac's bra-ket notation. 

%start with special case Z=C 
%change notation when comparing to qft operators
%specify time repr vs freq repr coherent state etc
The natural space to support the quantum noise, which models the effective action of the environment on the system, is a bosonic Fock space. It describes states of a quantum field (heat bath in our case) consisting of an indefinite number of identical particles.
The bosonic Fock space, over the one-particle space $\mathcal{H}$, is defined as 
\begin{equation} \Gamma(\mathcal{H}) =  \CC \oplus \mathcal{H} \oplus \mathcal{H}^{\circ 2} \oplus \dots \oplus \mathcal{H}^{\circ n} \oplus \dots, \end{equation} where $\CC$, denoting the one-dimensional space of complex numbers, is called the {\it vacuum subspace} and $\mathcal{H}^{\circ n}$, denoting the symmetric  product of $n$ copies of $\mathcal{H}$, is called the {\it $n$-particle subspace}. For any $n$ elements $|u_1\rangle, |u_2\rangle, \dots, |u_n\rangle$  in $\mathcal{H}$, the vector $\otimes_{j=1}^n |u_{j}\rangle$ is known as the Fock vector.  Since the particles constituting the noise space (and in each of the $n$-particle space) are bosons, in order to describe the $n$-particle state (i.e. to belong to the $n$-particle space, $\mathcal{H}^{\circ n}$), a Fock vector has to be symmetrized:   \begin{equation} |u_1\rangle \circ |u_2 \rangle \circ \dots \circ |u_n\rangle =  \frac{1}{n!} \sum_{\sigma \in \mathcal{P}_n} |u_{\sigma(1)}\rangle \otimes |u_{\sigma(2)}\rangle \otimes \dots \otimes |u_{\sigma(n)}\rangle, \end{equation}
where $\mathcal{P}_n$ is the set of all permutations $\sigma$, of the set $\{1,2,\dots,n\}$.

%A Fock vector in the $n$-particle subspace is $ \bigotimes_{j=0}^n u_j$, where the $u_j\in\mathcal{H}$ are a sequence of vectors. For bosons, this sequence has to be symmetric under the action of any element of the group of all permutations. 

Important elements of the bosonic Fock space, $\Gamma(\mathcal{H})$, are the {\it exponential vectors}: \begin{equation} |e(u)\rangle =  1 \oplus |u \rangle \oplus \frac{|u \rangle^{\otimes 2}}{\sqrt{2!}} \oplus \dots \oplus \frac{|u \rangle^{\otimes n}}{\sqrt{n!}} \oplus \dots, \end{equation}
where $|u \rangle \in \mathcal{H}$ and $|u\rangle^{\otimes n}$ denotes the tensor product of $n$ copies of $|u\rangle$.
% and is called the {\it $n$-particle vector}. 
The exponential vectors satisfy the following  scalar product formula:
\begin{equation} \langle e(u)|e(v) \rangle = e^{\langle u |  v \rangle} \end{equation} for every  $|u \rangle, |v \rangle \in \mathcal{H},$ with the same notation for scalar products in appropriate spaces.

The linear span, $\mathcal{E}$, of all exponential vectors forms a dense subspace of $\Gamma(\mathcal{H})$. We refer to $\mathcal{E}$ as the {\it exponential domain}. Any bounded linear operator on $\Gamma(\mathcal{H})$ can be determined by its action on the exponential vectors. Note that $|\psi(u)\rangle = e^{-\langle u| u \rangle/2} |e(u)\rangle$ is a unit vector.  The pure state with the density operator $|\psi(u) \rangle \langle \psi(u)|$ is called the {\it coherent state} associated with $|u\rangle$.  We call $|\Omega \rangle:= |e(0)\rangle$  the {\it Fock vacuum vector}, which corresponds to the state with no particles.  In the special case when $\mathcal{H} = \CC$, the coherent states on $\Gamma(\mathcal{H}) = \CC \oplus \CC \oplus \cdots$  are sequences of the form: \begin{equation}|\psi(\alpha)\rangle = e^{-|\alpha|^2/2} \left(1, \alpha, \frac{\alpha^2}{\sqrt{2!}}, \cdots, \frac{\alpha^{n}}{\sqrt{n!}} \dots \right).\end{equation}

%This can be viewed as a second quantized version of the coherent state \begin{equation}e^{-|\alpha|^2/2} \sum_{n=0}^{\infty} \frac{\alpha^{n}}{\sqrt{n!}} |n\rangle,\begin{equation} for any $\alpha \in \CC$, written as the expansion in the Fock basis (or occupation number basis) $\{|n\rangle\}_{n=0,1,\dots}$ with $|n\rangle$ representing the state with $n$ number of quanta (see, for instance, Chapter 4 of \cite{GardinerBook} or \cite{orszag2016quantum}).  Defining the usual  annihilation operator, $a$, and usual creation operator, $a^{\dagger}$, by their action: $a|0\rangle = 0$, $a|n\rangle = \sqrt{n} |n-1\rangle$ for $n=1,2,\dots$, $a^{\dagger}|n \rangle = \sqrt{n+1} |n+1\rangle$ for $n=0,1,\dots$, so that $|n\rangle = \frac{(a^{\dagger})^{n}}{\sqrt{n!}}|0\rangle $, we see that $a|\psi(\alpha)\rangle = \alpha |\psi(\alpha)\rangle,$ i.e.  the coherent state is an eigenstate of the usual annihilation operator. 

Consider now the case when the one-particle space is $\mathcal{H} = L^2(\RR^{+})$, leading to the bosonic Fock space $\Gamma(L^2(\RR^{+}))$. We will be formulating a differential (in time) description of processes on the Fock space, and  $\RR^+$ represents the time semi-axis. We emphasize that this has to be distinguished from the frequency representation, as adopted when writing the bath Hamiltonian $H_B$ in Section \ref{model}.
One can think of $\Gamma(L^2(\RR^{+}))$ as the space describing a single field channel coupled to the system. 

In general, one can consider the bosonic Fock space over the one-particle space, $L^2(\RR^+) \otimes \mathcal{Z} = L^2(\RR^+; \mathcal{Z})$, where $\mathcal{Z}$ is a complex separable Hilbert space, equipped with a complete orthonormal basis $(|z_{k}\rangle)_{ k \geq 1}$. The space $\mathcal{Z}$ is called the {\it multiplicity space} of the noise. An element of $L^2(\RR^+) \otimes \mathcal{Z}$ is a square integrable function from $\RR^{+}$ into $\mathcal{Z}$. Physically, the dimension of $\mathcal{Z}$ is the number of field channels coupled to the system. When $\mathcal{Z} = \CC$ (one-dimensional), the corresponding bosonic Fock space describes a single field channel \cite{nurdinlinear}. When $\mathcal{Z} = \CC^{d}$ and the $|z_{i}\rangle = (0, \dots , 0, 1, 0, \dots , 0)$ with $1$ in the $i$-th slot, $i = 1, 2,\dots, d$, is fixed
as a canonical orthonormal basis in $\CC^d$, the corresponding Fock space describes $d$ field channels coupled to the system. Since the dimension of $\mathcal{Z}$ can be infinite, it allows considering infinitely many field channels coupled to the system. To take advantage of this generality, we take the quantum noise space to be the bosonic Fock space $\Gamma(\mathcal{H})$ over $\mathcal{H} = L^2(\RR^+) \otimes \mathcal{Z}$ in the following. 

The canonical observables on the bosonic Fock space are the {\it creation} and {\it annihilation operators} associated to a vector $|u\rangle \in \mathcal{H}$, denoted $a^{\dagger}(u)$ and $a(u)$ respectively. They satisfy the commutation relations: $[a(u), a(v)] = 0$, $[a^{\dagger}(u), a^{\dagger}(v)] = 0$ and $[a(u), a^{\dagger}(v)] = \langle u | v \rangle = \int_{\RR^+} \overline{u(s)} v(s) ds$, for $|u\rangle, |v\rangle \in \mathcal{H}$.  Their action on the exponential vectors is defined by:
\begin{equation} a(u) |e(v)\rangle = \langle u | v \rangle |e(v)\rangle, \end{equation}
\begin{equation} a^{\dagger}(u) |e(v)\rangle = \frac{d}{d\epsilon} |e(v+\epsilon u)\rangle \bigg|_{\epsilon = 0}, \end{equation}
%\begin{equation} \langle e(v)|a(u)|e(w) \rangle = \langle a^{\dagger}(u) e(v) | e(w) \rangle,\begin{equation} 
for all $|u\rangle, |v\rangle \in \mathcal{H}$. Note that in the special case $u=v$, we have $a(u) |\psi(u)\rangle = \langle u | u \rangle |\psi(u)\rangle$, which is an eigenvalue relation similar to the one that defines the coherent state as eigenvector of annihilation operator in quantum optics \cite{glauber1963coherent}. Moreover, we have
\begin{align} a(u) |e(0)\rangle &= 0, \\ 
a(u) |v\rangle^{\otimes n} &= \sqrt{n} \langle u | v \rangle |v\rangle^{\otimes (n-1)}, \\ 
a^{\dagger}(u)  |v\rangle^{\otimes n} &= \frac{1}{\sqrt{n+1}} \sum_{r = 0}^{n} |v\rangle^{\otimes r } \otimes |u\rangle \otimes |v\rangle^{\otimes (n-r)}.\end{align} Since vectors of the form $|v\rangle^{\otimes n}$ linearly span the $n$-particle space, this shows that $a(u)$ maps the $n$-particle subspace into the $(n-1)$-particle subspace while $a^{\dagger}(u)$ maps the $n$-particle subspace into the $(n+1)$-particle subspace, justifying their names as annihilation and creation operators respectively. 

The basic idea of H-P quantum stochastic calculus comes from the {\it continuous tensor product factorization} property of bosonic Fock space. The tensor product factorization property says that when the one-particle space is given by a direct sum, $\mathcal{H} = \mathcal{H}_1 \oplus \mathcal{H}_2$,  we have the factorization property for the corresponding Fock space: $\Gamma(\mathcal{H}) = \Gamma( \mathcal{H}_1 \oplus \mathcal{H}_2) = \Gamma( \mathcal{H}_1) \otimes \Gamma(\mathcal{H}_2)$. In our setup, 
\begin{equation} 
L^2(\RR^{+}; \mathcal{Z}) = L^2([0,t];\mathcal{Z}) \oplus L^2([t,\infty);\mathcal{Z}) \end{equation}
for every $t > 0$. We exploit this property to describe the total space on which the system and the noise evolve jointly. Denote, for all $0 < s<t$:
\begin{equation} \mathcal{F}_{t]}= \mathcal{H}_{S} \otimes \Gamma(L^2([0,t];\mathcal{Z})), \ \ \mathcal{F}_{[s,t]} = \Gamma(L^2([s,t];\mathcal{Z})), \ \ \mathcal{F}_{[t} = \Gamma(L^2([t,\infty);\mathcal{Z})), \end{equation}
with $\mathcal{F}_{0]} = \mathcal{H}_{S}$ (the initial space describing the system) and $\mathcal{F} = \mathcal{H}_{S} \otimes \Gamma(\mathcal{H})$ (the total space for the time evolution of the system in the presence of quantum noise).
We then have the natural identification  $\mathcal{F} = \mathcal{F}_{t]} \otimes \mathcal{F}_{[t}$ based on the factorization of the exponential vectors: \begin{equation} |\psi \rangle \otimes |e(u) \rangle =  |\psi\rangle \otimes |e(u_{[0,t]}) \rangle \otimes |e(u_{[t,\infty)}) 
\rangle,\end{equation} where $|\psi\rangle \in \mathcal{H}_{S}$, $u_{[0,t]}(\tau) = 1_{[0,t]}(\tau) u (\tau)$, $u_{[t,\infty)}(\tau) = 1_{[t,\infty)}(\tau) u (\tau)$. Note that $\mathcal{F}_{t]}$ and $\mathcal{F}_{[t}$ embed naturally into $\mathcal{F}$ as subspaces by tensoring with the vacuum vector. 

%$\mathcal{Z}$-valued 
Any vector $|u\rangle \in \mathcal{H}$ may be regarded as a $\mathcal{Z}$-valued function.  For a fixed basis of $\mathcal{Z}$ (e.g. in the case when $\mathcal{Z}$ is the space  $\CC^d$ with the canonical basis $|z_k\rangle$), we set  $u_k(t)  = \langle z_k | u(t) \rangle_{\mathcal{Z}}$ for $k \geq 1$, where $\langle \cdot | \cdot \rangle_{\mathcal{Z}}$ denotes scalar product on $\mathcal{Z}$.  We define the {\it creation and annihilation processes} associated with the orthonormal basis $\{|z_{k}\rangle \}_{k\geq 1}$ as follows:
\begin{equation} \label{cr_ann} A^{\dagger}_{k}(t) = a^{\dagger}(1_{[0,t]} \otimes z_{k}), \ \ \ \ \  \ A_{k}(t) = a(1_{[0,t]} \otimes z_{k}),\end{equation} for $k=1,2,\dots$, where $1_{[0,t]}$ denotes indicator function of $[0,t]$ as an element of $L^2(\RR^{+})$. Each $A_{k}$ (respectively, $A_{k}^{\dagger}$) is defined on a distinct copy of the Fock space $\Gamma(L^2(\RR^{+}))$ and therefore, the $A_{k}$'s (respectively, $A_k^{\dagger}$) are commuting. Physically, each of them represents a  single channel of quantum noise input coupled to the system. Note that in the special case $\mathcal{Z} = \CC$, the above construction only gives a single pair of creation and annihilation process and in the case $\mathcal{Z} = \CC^{d}$, we have $d$ pairs of creation and annihilation processes associated with $d$ distinct noise inputs. The actions of the $A_k(t)$ on the exponential vectors are given by the eigenvalue relations: 
\begin{equation} \label{reg1}
A_k(t)|e(u)\rangle = \left(\int_0^t u_k(s) ds \right) |e(u)\rangle, 
\end{equation}
and the $A_k^{\dagger}(t)$ are the  corresponding adjoint processes:
\begin{equation} \label{reg2}
\langle e(v) | A_k^{\dagger}(t) | e(u) \rangle = \left(\int_0^t \overline{v_k(s)} ds \right) \langle e(v) | e(u) \rangle. \end{equation}
The above processes, which are time integrated versions of instantaneous creation and annihilation operators are two of the three kinds of {\it fundamental noise processes} introduced by Hudson and Parthasarathy.
They satisfy an integrated version of the CCR: $[A_k(t), A_l^{\dagger}(s)] = \delta_{kl} \mbox{min}(t,s) $, $[A_k(t), A_l(s)] = [A_k^{\dagger}(t), A_l^{\dagger}(s)] = 0$. 

For each $k$, their `future pointing' infinitesimal time increments, $dA^{\#}_k(t):=A_k^{\#}(t+dt) - A^{\#}_k(t)$, where $\#$ denotes either creation or annihilation processes, with respect to the time interval $[t,t+dt]$, are {\it independent} processes. The independence is due to the fact that time increments with respect to non-overlapping time intervals are commuting since they are {\it adapted} with respect to $\mathcal{F}$, i.e. they act non-trivially on the factor $\mathcal{F}_{[t,t+dt]}$ of the space $\mathcal{F} =\mathcal{F}_{t]}\otimes \mathcal{F}_{[t,t+dt]} \otimes \mathcal{F}_{[t+dt}$ and trivially, as identity operator on the remaining two factors. In other words, for a fixed $k$, 
\begin{align}
dA_{k}^{\#}(t) | e(u) \rangle &= (A_k^{\#}(t+dt) - A^{\#}_k(t)) |e(u) \rangle \\ 
&= e(u_{[0,t]}) \otimes  a^{\#}(1_{[t,t+dt]} \otimes z_k)  e(u_{[t,t+dt]}) \otimes  e(u_{[t+dt,\infty)}), 
\end{align}
where the operators $a^{\#}$ are defined in \eqref{cr_ann}. Therefore, any Hermitian noise processes $M(t)$ that are appropriate combinations of the $A_k^{\#}(t)$ (for instance, the quantum Wiener processes introduced later in \eqref{qwp}) have independent time increments, i.e. if we define the characteristic function of $M$ with respect to the coherent state, $|\psi(u) \rangle$, as $\varphi_{M}(\lambda) := \langle \psi(u)| e^{i\lambda M} |\psi(u)\rangle$, then for any two times $s \leq t$, we see that their joint characteristic function with respect to the coherent states is the product of individual characteristic functions: 
\begin{align}
\varphi_{M(s),M(t)-M(s)}(\lambda_s,\lambda_t) &:= \langle \psi(u)| e^{i\lambda_s M(s) + i \lambda_t (M(t)-M(s))} |\psi(u)\rangle \nonumber \\ 
&= \varphi_{M(s)}(\lambda_s) \varphi_{M(t)-M(s)}(\lambda_t). 
\end{align}
This property is a quantum analog of the notion of processes with independent increments in classical probability. 

%In the special case $\mathcal{Z} = \CC^d$, which corresponds to the case when the noise comes from $d$ field channels,  we have $d$ independent annihilation and creation processes, each associated to a distinct field channel and is defined on a distinct copy of the Fock space. 

\begin{rmk} \label{qft}
In quantum field theory, the operators  $A^{\dagger}_{k}(t)$ and $A_{k}(t)$ are called the smeared field operators and are usually written formally as:
\begin{equation} A_k(t) = \int_0^t b_k(s) ds, \ \ \ \  \ \ A_{k}^{\dagger}(t) = \int_0^t b^{\dagger}_k(s) ds, \end{equation}
where the $b_k(t) = \frac{1}{\sqrt{2 \pi}}\int_{\RR} \hat{b}_k(\omega) e^{-i\omega t} d\omega$ and $b_k^{\dagger}(t)=\frac{1}{\sqrt{2 \pi}}\int_{\RR} \hat{b}^{\dagger}_k(\omega) e^{i\omega t} d\omega$ are the idealized Bose field processes  satisfying the singular CCR: $[b_k(t), b^{\dagger}_l(s)] = \delta_{kl} \delta(t-s)$ \cite{GardinerBook}. Physically, since $b^{\dagger}_{k}(s)$ creates a particle at time $s$ through the $k$th noise channel, $A_{k}^{\dagger}(t)$ creates a particle that survives up to time $t$. These formal expressions for the annihilation and creation processes are simpler to work with than the more regular integrated processes defined in \eqref{reg1}-\eqref{reg2}. As remarked on page 39 of \cite{nurdinlinear}, the more fundamental processes from the underlying
physics point of view are the quantum field  processes, not the rigorously defined, more regular integrated processes.  \end{rmk}

Exploiting the structure of bosonic Fock space and the properties of the fundamental noise processes outlined above, Hudson and Parthasarathy developed and studied {\it quantum stochastic integrals} with respect to these fundamental processes  for a suitable class of adapted integrand processes, in analogy with the constructions of the classical It\^o theory. The most important result of the calculus is the {\it quantum It\^o formula}, which describes how the classical Leibnitz formula for the time-differential of a product of two functions gets corrected when these functions depend explicitly on the fundamental processes. In the vacuum state, the quantum It\^o formula can be summarized by:
\begin{equation} \label{qif}
dA_k(t) dA^{\dagger}_l(t) = \delta_{kl} dt \end{equation} and all other products of differentials that involve $dA_k(t)$, $dA_k^{\dagger}(t)$ and $dt$ vanish. This can be viewed as a chain rule with Wick ordering \cite{streater2000classical} and as a quantum analogue of the classical It\^o formula. 

In particular, with respect to the initial vacuum state, the field quadratures $W^0_k(t) = A_k(t) +A_k^{\dagger}(t)$ ($k=1,2,\dots$) are mean zero Hermitian Gaussian processes with variance $t$. Therefore, they can be viewed as quantum analogue of classical Wiener processes and their formal time derivatives, $dW^0_k(t)/dt = b_k(t) + b^{\dagger}_k(t)$, are quantum analogues of the classical white noises. If one takes $\mathcal{Z} = \CC^{d}$, then $(W^0_1,W^0_2,\dots,W^0_{d})$ is a collection of commuting processes and thus form a quantum analogue of $d$-dimensional classical Wiener process in the vacuum state. Moreover, one has $dW^0_i(t) dW^0_j(t) = \delta_{ij}dt$, which is the classical It\^o correction formula for Wiener process. These results hold for a more general class of field observables: 
\begin{equation} \label{qwp}
W^\theta_k(t) = e^{-i\theta_k } A_k(t) + e^{i\theta_k} A^{\dagger}_k(t),
\end{equation}
where $\theta_k \in \RR$ is a phase angle. We refer to the $W^\theta_k(t)$ as {\it quantum Wiener processes}, as they are operator-valued processes on a Fock space, analogous to classical Wiener processes.
%gauge freedom

In contrast to closed quantum system's unitary evolution, interaction with an environment leads to randomness in the unitary evolution of an open quantum system. Using the quantum It\^o formula, Hudson and Parthasarathy deduced the general form of a unitary, reversible, Markovian evolution for a system interacting with an environment described by the fundamental noise processes. The unitary evolution operator, $V(t)$, of the whole system, in the interaction picture with respect to the free field dynamics, is found to satisfy an It\^o SDE of the following form:
\begin{align}
dV(t) &= \left[ \left( -\frac{i}{\hbar} H_e -  \frac{1}{2} \sum_{k} L^{\dagger}_k L_k \right) dt + \sum_{k} \left( L_k^{\dagger} dA_k(t) - L_k dA_k^{\dagger}(t) \right) \right] V(t), \\
V(0) &= I, 
\end{align}
associated to the system operators $(H_e, \{L_k\})$, where $H_e = H_e^{\dagger}$ is an effective Hamiltonian and the $L_k$ are Lindblad coupling operators. It can be viewed as a noisy Schrodinger equation. The evolution of a noisy system observable, $X$, initially defined on $\mathcal{H}_{S}$, can also be obtained. By applying the quantum It\^o formula, one can deduce that its evolution, $j_t(X) = V(t)^{\dagger} (X \otimes I) V(t)$, on $\mathcal{F}$ is described by the following Heisenberg-Langevin equation:
\begin{align}
dj_t(X) &= j_t(\mathcal{L}(X))dt + \sum_k \bigg( j_t([X,L_k])dA_k^{\dagger}(t) + j_t([L_k^{\dagger},X])dA_k(t) \bigg), \\ 
j_{0}(X) &= X \otimes I, \end{align}
where $\mathcal{L}$ is the well-known Lindblad generator \cite{Lindblad1976}: 
\begin{equation}
\mathcal{L}(X) = \frac{i}{\hbar}[H_e,X] + \frac{1}{2} \sum_k ([L_k^{\dagger},X]L_k + L_k^{\dagger} [X,L_k] ). \end{equation} One can also obtain the evolution of field observables in  this way and study the relation between input and output field processes \cite{GardinerBook}. We call such equation for an observable a {\it quantum stochastic differential equation} (QSDE) and its solution is a {\it quantum stochastic process}, which is a noncommutative analogue of classical stochastic process. To obtain the Lindblad master equation for reduced system density operator, $\rho_S(t)$, we first take the Fock vacuum conditional expectation of $j_t(X)$ to obtain the evolution of the reduced system observable, $T_t(X)$, defined via \begin{equation} \langle \psi| T_t(X) | \phi \rangle = \langle \psi \otimes e(u) | j_t(X) | \phi \otimes e(v) \rangle, \end{equation} so that $dT_{t}(X) = T_{t}(\mathcal{L}(X)) dt$,  then the Lindblad master equation:
\begin{equation} d\rho_S(t) = \mathcal{L}^{*}(\rho_{S}(t)) dt\end{equation} is obtained by duality. 

%correl func/char func 
\begin{rmk} Following Remark \ref{qft}, one can introduce the notion of {\it quantum colored noise} \cite{belavkin1995world,xue2017modelling}. For $k=1,2,\dots$, define \begin{equation} b^{\dagger}_{g,k}(t) := \frac{1}{\sqrt{2 \pi}} \int_{\RR} \hat{b}_k^{\dagger}(\omega) e^{i\omega t} \hat{g}(\omega) d\omega, \ \ \ \ \ \  \ b_{g,k}(t) := \frac{1}{\sqrt{2 \pi}} \int_{\RR} \hat{b}_k(\omega) e^{-i\omega t} \overline{\hat{g}(\omega)} d\omega,\end{equation}
where $\hat{g}(\omega)$ and $\hat{b}_k(\omega)$ denote the Fourier transform of $g(t)$ and $b_k(t)$ respectively. Note that $b^{\dagger}_{g,k}(t)$ is the inverse Fourier transform of $\hat{b}_k^{\dagger}(\omega) \hat{g}(\omega)$ and so by the convolution theorem we have: \begin{equation} b^{\dagger}_{g,k}(t) = \frac{1}{\sqrt{2 \pi}} \int_{\RR} g(t-s) b_k^{\dagger}(s) ds = \frac{1}{\sqrt{2 \pi}} \int_{\RR} g(t-s) dA_k^{\dagger}(s),\end{equation} where the $A_{k}^{\dagger}(s)$ are creation processes. This is reminiscent of the formula for classical colored noise defined via filtering of white noise \cite{lindgren2006lectures}: 
\begin{equation} \int_{\RR} \gamma(t-s)1_{\{s\leq t \}}(s) dB_{s} = \int_{-\infty}^t \gamma(t-s) dB_{s},  \end{equation} where $\gamma(t)$ describes the filter and $B=(B_s)$ is a classical Wiener process. In the limit $g \to 1$ (flat spectrum limit), the $b^{\dagger}_{g,k}(t)$  converge to the fundamental noise process $b_k^{\dagger}(t) = dA_k^{\dagger}/dt$. Similar remarks apply to $b_{g,k}(t)$ and to appropriate linear combinations of  $b_{g,k}(t)$ and  $b^{\dagger}_{g,k}(t)$, which therefore deserve to be called quantum colored noise processes. \end{rmk}

%-----------------------------------------

\section{QSDE's for Quantum Noise} \label{qnoise}

Guided by formula of the symmetric correlation function in \eqref{D1} and the plan outlined in Section \ref{approach}, we model the quantum noise by: 
\begin{equation}
\label{eq:noise}
 \sum_{k=0}^{\infty} \eta_{k}(t),
\end{equation}
  where the $\eta_{k}(t)$ are independent {\it quantum Ornstein-Uhlenbeck processes} \cite{csaki1991infinite}, satisfying the SDEs: 
\begin{equation} \label{qou} d\eta_k(t) = -\alpha_{k} \eta_{k}(t) dt + \sqrt{\lambda_{k}} dW^\theta_{k}(t), \ \ \eta_k(0) = \eta_k.
\end{equation}
Here the $W_k^{\theta}$ are independent quantum Wiener processes defined in Section \ref{background_qsde} and the $\eta_{k}$ are  initial variables on a copy of Fock space. For a fixed $\theta$, independence and commutation for these processes can be achieved by realizing  the $\eta_{k}(t)$ on distinct copies of Fock space, i.e. \begin{equation} \sum_{k=0}^{\infty} \eta_k(t) = \eta_{0}(t) \otimes I \otimes I \otimes \dots + I \otimes \eta_1(t) \otimes I \otimes \dots + \dots  \end{equation} on $\bigotimes_{k=0}^{\infty} \Gamma(L^2(\RR^{+})) = \Gamma(L^2(\RR^+)\otimes \mathcal{K})$ where the multiplicity space $\mathcal{K}$ is a sequence space whose elements are of the form $(x_0, x_1, x_2, \dots)$, with each $x_i \in \CC$. From now on, each $\eta_k$ is understood to be \begin{equation} \underbrace{I \otimes \cdots \otimes I}_{k \text{  copies}} \otimes \eta_k \otimes I \otimes \cdots \end{equation} and similarly for each $W_{k}^{\theta}$. 

The formal solution to the SDE \eqref{qou} is given by:
\begin{equation} \eta_{k}(t) = \eta_k e^{-\alpha_{k} t} + \sqrt{\lambda_{k}} \int_{0}^{t} e^{-\alpha_{k}(t-s)} dW^{\theta}_{k}(s). \end{equation} 

Since there is a unique stationary solution of the SDEs \eqref{qou}, for all $k$ and $s \in [0,t]$:
\begin{equation} E''_{\infty}[\eta_{k}^2] = \frac{\lambda_k}{2 \alpha_k}, \ \ \ \ E''_\infty[\eta_k W^\theta_{k}(s)] = E''_{\infty}[W^\theta_k(s) \eta_k] =  0,\end{equation} where  $E''_{\infty}$ denotes expectation with respect to the vacuum state associated with $\Omega \otimes \Omega \otimes \cdots$ on the enlarged Fock space $\Gamma(L^2(\RR^{+})\otimes \mathcal{K})$. Then, with the parameters $\alpha_n$  and $\lambda_{n}$ defined by
\begin{equation}
\label{eq:alphan}
\alpha_{n} = \nu_n 1_{\{n \geq   1 \}} + \Lambda 1_{\{n=0\}} > 0 
\end{equation} and
\begin{equation}
\label{eq:lambdan}
\lambda_{n} = \frac{4 \nu_n^2 \Lambda^2 k_B T}{\nu_n^2-\Lambda^2}  1_{\{n \geq 1 \}} + \hbar \Lambda^3 \cot\left( \frac{\hbar \Lambda}{2k_{B}T} \right)  1_{\{n=0\}} > 0,
\end{equation} 
it can be verified that 
\begin{equation}
E''_{\infty}\left[\frac{\left\{\sum_k \eta_k(t), \sum_l \eta_{l}(s) \right\}}{2} \right] = D_{1}(t-s),
\end{equation} 
where $D_{1}$ is given in \eqref{D1}. 

Equations~\eqref{eq:alphan} and~\eqref{eq:lambdan} establish a link between the quantum noise as introduced in eqn.~\eqref{eq:noise} and the physical model of Section \ref{model}. We remark that there is freedom in the above construction of quantum noise, as the driving noise process, $(W_k^{\theta})$, is a family of quantum Wiener processes parametrized by $\theta$. On the one hand, the choice of the parameter should be fixed by physical considerations, i.e. by the nature of the field that the system couples to in the microscopic model. On the other hand, one would like to show that the quantum noises describe a Markovian system, so one should write the SDEs $\eqref{qou}$ in a H-P QSDE form.

To this end, let $\xi_{k}(t)$ and $\eta_{k}(t)$ be canonical conjugate bath observables that obey the commutation relation $[\xi_{j}(t), \eta_{k}(t) ] = i \hbar \delta_{jk} I$ for all $t \geq 0$. Suppose that the evolution of each pair $(\xi_k(t), \eta_{k}(t))$ is Markovian and can be described by the H-P QSDEs associated with $(H_k, L_k)$, where 
\begin{equation} H_{k} = \frac{\eta_k^2}{2} + \frac{\alpha_{k}}{4} \{ \xi_k, \eta_k \}, \ \  \ \ L_{k} = \frac{\sqrt{\lambda_k}}{\hbar} \xi_k + i \frac{\alpha_{k}}{2\sqrt{\lambda_{k}}} \eta_{k},\end{equation} where $\alpha_{k}$ and $\lambda_{k}$ are given as before. Therefore, they solve the  H-P QSDEs:
\begin{align}
d\xi_k(t) &= \eta_k(t) dt + \frac{\hbar \alpha_k}{2 \sqrt{\lambda_k}} dW^\pi_{k}(t), \label{qn1} \\
d \eta_{k}(t) &= -\alpha_{k} \eta_{k}(t) dt + \sqrt{\lambda_{k}} dW^{-\pi/2}_{k}(t), \label{qn2}
\end{align}
where \begin{equation} W^\pi_{k}(t) = -(A_k(t) + A^\dagger_k(t)), \ \ \ W^{-\pi/2}_{k}(t) = i(A_k(t) - A^\dagger_k(t))\end{equation} are noncommuting, conjugate quantum Wiener processes satisfying $[W_k^{\pi}(t), W_{k}^{-\pi/2}(s)] = 2i \delta(t-s) I$.  Modulo the negative factor, one can view the formal time derivatives of the $W^\pi_{k}(t)$ and $W^{-\pi/2}_{k}(t)$ as the noises arising from the position and momentum field observables respectively. We fix the freedom in our construction by taking the  Markovian system \eqref{qn1}-\eqref{qn2} as the model for noise. Therefore, we take $\sum_k \eta_k(t)$ to be the quantum colored noise that models the action of the heat bath on the evolution of the system's observables. 
% In other words, we have modeled the whole system by the following (self-adjoint) operator on $\mathcal{H}_S \otimes \Gamma(L^2(\RR^+)\otimes \mathcal{K})$, which in the interaction picture with respect to the free field, reads: \begin{equation}H(t) = (H_{S}+H_{ren}) \otimes I - f(X) \otimes  \sum_{k} \left( \eta_k e^{-\alpha_{k}t} + \sqrt{\lambda_{k}}  \int_{0}^{t} e^{-\alpha_{k}(t-s)} dW^{-\pi/2}_{k}(s) \right).\begin{equation}

Physically, one can think of our quantum noise model as equivalent to a model of infinitely many non-interacting ancillas that convert the white noise to colored noise through a channel at each Matsubara frequency \cite{xue2015quantum}. That one needs  infinitely many ancillas is due to the fact that there are infinitely many transition (Bohr) energies, each of which equals  the energy of a boson with a particular Matsubara frequency in the bath. According to our noise model, when a boson with the Matsubara frequency $\nu_{k}$ is created or annihilated, the energy transition does not occur instantaneously but happens on the time scale of $1/\alpha_k$ via a channel associated with $\nu_k$.

%Note that this distinguishes from the classical situation, where there is no noise in the equation for the "position" processes $\xi(t)^{(k)}$. We will see, however, that such position processes do not survive in the $\epsilon \to 0$ limit of our rescaled model.

%\rmk The modeling of the quantum noises as quantum stochastic processes that satisfy certain QSDEs driven by H-P fundamental noise processes is crucial to derive quantum noise-induced drifts in next section (i.e. we need the presence of quantum analogues of classical white noise in the Langevin equation  in order to obtain correction drifts that are really induced by the white noise in the limiting equation). It is this line of thinking that makes Langevin approach particularly well suited for the derivation of noise-induced drifts, as was done for the classical case. However, this comes at the cost of having to deal with the singular nature of the unbounded operator-valued Langevin equations. There are also some subtleties and pitfalls  when adopting the Langevin approach. 

%------------------------------------------

\section{The Rescaled Model and SDE's}
\label{rescaledmodel}
We set $m = m_{0} \epsilon$, $\Lambda = E_{\Lambda}/\epsilon$ and $\hbar = \epsilon$, where $m_{0}$ and $E_{\Lambda}$ are fixed positive constants and $\epsilon > 0$ is a small parameter, so that $\hbar \Lambda = E_{\Lambda}$ (the maximum energy of bosons in the bath) and $m/\hbar = m_{0}$ (proportional to the de Broglie wavelength of the Brownian particle)  are fixed in the Hamiltonian.

These scalings of the model parameters are motivated as follows. An atom's mass is often small and so are the characteristic time scales of the quantum bath. On the other hand, for small Planck constant the equipartition theorem is valid, so the mean kinetic energy of the system is O(1) (i.e. of order $1$) as $\epsilon \to 0$. This has to be compared to the classical case, where the fact that the kinetic energy is  O(1) leads to the presence of noise-induced drift in the small $m$ limit when the original system is subject to state-dependent damping and diffusion \cite{volpe2016effective}. Hence, this suggests that the scalings give meaningful effective dynamics in the limit $\epsilon \to 0$. 

%After the scalings, we have the rescaled noise kernel: \begin{equation} D_{1}^{\epsilon}(t) = \frac{E_{\Lambda}^2}{2\epsilon} \cot \left(\frac{E_{\Lambda}}{2k_{B}T} \right) e^{-\frac{E_{\Lambda}}{\epsilon} |t|} +  \sum_{n=1}^{\infty} \frac{2 E_{\Lambda}^2 k_{B}T }{(2 \pi n k_{B}T)^2-E_{\Lambda}^2 } \frac{2 \pi n k_{B}T  }{ \epsilon} e^{-\frac{2 \pi n k_{B}T}{ \epsilon}|t| }, \begin{equation} and the rescaled dissipation kernel is \begin{equation}D^{\epsilon}(t) = \frac{E_{\Lambda}^2}{2 \epsilon} e^{-\frac{E_{\Lambda}}{\epsilon} |t|}  = \frac{E^2_{\Lambda}}{2} \frac{1}{\epsilon} \kappa(|t|/\epsilon) .\begin{equation}

Next, we elucidate our scalings  in the context of separation of time scales. Taking $\epsilon \to 0$ is equivalent to taking the joint limit of small mass ($m \to 0$), the memoryless limit ($\Lambda \to \infty$) (which also implies the small noise correlation time limit, due to the quantum fluctuation-dissipation relation) and the classical limit ($\hbar \to 0$). Note that in the limit the spectral density $J(\omega)$ becomes strictly Ohmic, since the cutoff is removed as $\epsilon \to 0$. In other words, the inertial time scale, the memory time scale, the noise correlation time scale and the quantum time scale vanish simultaneously at the same rate as we take $\epsilon \to 0$ in the rescaled model. This limit is a quantum version of the one studied in \cite{LimWehr_Homog_NonMarkovian}.

Upon applying the above rescalings, the parameters in the QSDEs for the $\xi_{n}(t)$ and $\eta_n(t)$ in Section \ref{qnoise} become
\begin{equation}
\label{eq:a_n}
\alpha_{n} = \frac{2 \pi n k_{B}T}{\epsilon} 1_{\{n \geq 1 \}} + \frac{E_{\Lambda}}{\epsilon} 1_{\{n=0\}} =: \frac{1}{\epsilon} a_n,
\end{equation}
and
\begin{equation}
\label{eq:E_n}
 \lambda_{n} = \frac{4 k_{B}T (2 \pi n k_{B}T)^2}{(2 \pi n k_{B}T)^2 - E_{\Lambda}^2} \frac{E_\Lambda^2}{\epsilon^2} 1_{\{n \geq 1\}} + E_{\Lambda} \cot\left( \frac{E_{\Lambda}}{2k_{B}T} \right) \frac{E_{\Lambda}^2}{\epsilon^2} 1_{\{n=0\}} =: \frac{1}{\epsilon^2} \Sigma_n^2.
 \end{equation}
 Notice that in the following the relevant parameters are $a_n$ and $\Sigma_n$ as defined in eqns.~\eqref{eq:a_n} and~\eqref{eq:E_n}. The rescaled version of the resulting Heisenberg-Langevin equations \eqref{qle1}-\eqref{qle2} can be cast as the following system of SDEs on the total space $\mathcal{F} = \mathcal{H}_S \otimes \Gamma(L^2(\RR^+) \otimes \mathcal{K})$: 
\begin{align}
dX(t) &= V(t)  dt, \label{res_hl1} \\  
m_{0} \epsilon d V(t) &= -U'(X(t)) dt + f'(X(t)) \sum_{n=0}^{\infty} \eta_{n}(t)  dt - f'(X(t)) Y(t) dt, \\ 
dZ(t)  &= Y(t)  dt, \\ 
\epsilon dY(t)   &= -E_{\Lambda} Y(t) dt + E_{\Lambda} f'(X(t)) V(t) dt - i \frac{E_{\Lambda}}{2m_{0}} f''(X(t)) dt, \label{Yeq} \\ 
d \xi_{n}(t) &= \eta_{n}(t) dt + \epsilon \frac{a_n}{ 2 \Sigma_n}  dW_n^{\pi}(t), \ n = 0,1,2, \dots, \\
\epsilon d\eta_{n}(t) &= - a_{n} \eta_{n}(t) dt+ \Sigma_n dW_n^{-\pi/2}(t), \ \ n = 0,1,2, \dots,  \label{res_hl6}
\end{align}
where we have defined the auxiliary quantum stochastic process \begin{equation} Y(t) = \frac{E_{\Lambda}}{\epsilon} \int_{0}^{t} e^{-\frac{E_{\Lambda}}{\epsilon}(t-s)} \frac{\{f'(X(s)), V(s) \}}{2} ds\end{equation} and used the commutation relation $\frac{[P,f'(X)]}{2m} = -\frac{i f''(X)}{2m_{0}}$ to rearrange the order, so that $V(t)$ appears last in \eqref{Yeq}. Note that it is crucial that we have the scaling $\frac{\hbar}{m} = \frac{1}{m_{0}}$ so that the last term on the right hand side of \eqref{Yeq} is $O(1)$. It should be clear from the context which factor of the total space the operators act non-trivially on. For instance, $X = X(t=0) = X(t=0) \otimes I_0 \otimes I_1 \otimes \dots $; $\xi_n(t) = I \otimes I_0 \otimes \dots \otimes I_{n-1} \otimes \xi_n(t) \otimes I_{n+1} \otimes \dots ,$ for $n=0,1,\dots $; etc, where $I$ is identity operator on $\mathcal{H}_{S}$ and $I_{n}$ is identity operator on the $n$th copy of Fock space. 

From now on, vectors and matrices whose elements are operators will be denoted by bold letters. For an operator matrix $\vecc{A} = (A_{ij})_{i,j=0,1,2,\dots}$, its transpose, denoted by $^T$, is defined as $(A_{ij})^{T}_{i,j=0,1,2,\dots} := (A_{ji})_{i,j=0,1,2,\dots}.$ The action of $\vecc{A}$ on an operator vector $\vecc{x}= (x_j)_{j=0,1,2,\dots}$, written as $\vecc{A} \vecc{x}$, results in another operator vector $(\sum_j A_{ij}x_j)_{i=0,1,2,\dots}$.  If $\vecc{A}$ is diagonal, we write it as $\mbox{diag}(A_k)_{k=0,1,2,\dots}$. 

Introducing the operator vectors
\begin{equation} \vecc{X}_{t} = [X(t)   \ \ Z(t)   \ \ \xi_{0}(t)  \ \cdots \ \xi_{N}(t) \ \ \cdots]^{T},\ \ \ \ \vecc{V}_{t} = [V(t)  \ \ Y(t)  \ \ \eta_{0}(t) \ \cdots \ \eta_{N}(t)  \ \ \cdots]^{T}, \end{equation}
we rewrite the above system in a more compact way:
\begin{align}
d\vecc{X}_{t} &= \vecc{V}_{t}dt + \epsilon \vecc{\mu} d\vecc{W}^{\pi}_t, \label{x_eqn} \\ 
\epsilon d\vecc{V}_{t} &= - \vecc{\hat{\gamma}}(X(t)) \vecc{V}_{t} dt + \vecc{F}(X(t)) dt +  \vecc{\sigma} d\vecc{W}^{-\pi/2}_{t}, \label{v_eqn}
\end{align}
with the initial conditions $\vecc{X}_t = \vecc{X}$ and $\vecc{V}_t = \vecc{V}$. 

In the above, $\vecc{\hat{\gamma}}(X(t))$ (the superoperator that acts on $\vecc{V}_t$) denotes the block operator matrix, whose entries depend on $X(t)$, given by \begin{equation}
 \vecc{\hat{\gamma}}(X(t)) =  \left[ \begin{array}{cc}
\vecc{A}(X(t)) & \vecc{B}(X(t)) \\
\vecc{0} & \vecc{D} 
\end{array} \right],  \end{equation}
with
\begin{align} \vecc{A}(X(t)) &=  \left[ \begin{array}{cc}
0 & \frac{f'(X(t))}{m_{0}} \\
-E_{\Lambda} f'(X(t)) & E_{\Lambda}
\end{array} \right], \\ 
\vecc{B}(X(t)) &=  \left[ \begin{array}{ccc}
-\frac{f'(X(t))}{m_{0}}  & -\frac{f'(X(t))}{m_{0}} & \cdots\\
0 & 0 & \cdots
\end{array} \right], \ \ \ \ \ \  \vecc{0} =  \left[ \begin{array}{cc} 0 & 0  \\
0 & 0 \\ \vdots & \vdots \end{array} \right],  \end{align}
 
$\vecc{D} = \mbox{diag}(a_n)_{n=0,1,\dots}$ is the diagonal operator matrix,
\begin{equation} \vecc{F}(X(t)) = \left[-\frac{U'(X(t))}{m_{0}}  \ \ -i\frac{E_{\Lambda}}{2m_{0}} f''(X(t))  \ \ 0 \  \ 0 \ \ \cdots   \right]^{T} , \end{equation}
%\[ \vecc{\Sigma} =  \left[ \begin{array}{cccc}0 & 0 & \dots & 0   \\ 0 & 0 & \dots & 0 \\ \frac{1}{\sqrt{E_{\Lambda}    \cot\left(\frac{E_{\Lambda}}{2k_{B}T} \right)}} \psi  & 0 & \dots  & 0 \\  0 & \sqrt{\frac{(2e  k_{B}T)^2-E_{\Lambda}^2}{4k_{B}TE_{\Lambda}^2} } \psi & 0  & 0 \\  \vdots & \vdots & \ddots & \vdots \\  0 & 0 & 0 & \sqrt{\frac{(2e N k_{B}T)^2-E_{\Lambda}^2}{4k_{B}TE_{\Lambda}^2} } \psi \end{array} \right],  \]

%\[ \vecc{\sigma} =  \left[ \begin{array}{cccc} 0 & 0 & \dots & 0   \\ 0 & 0 & \dots & 0 \\ E_{\Lambda} \sqrt{E_{\Lambda} \cot\left(\frac{E_{\Lambda}}{2k_{B}T} \right)} \psi & 0 & \dots  & 0 \\  0 & \frac{4 e  (k_{B}T)^{3/2} E_{\Lambda}}{\sqrt{(2 \pi  k_{B}T)^2 - E_{\Lambda}^2 }} \psi & 0  & 0 \\  \vdots & \vdots & \ddots & \vdots \\  0 & 0 & 0 & \frac{4 \pi N (k_{B}T)^{3/2} E_{\Lambda}}{\sqrt{(2e N k_{B}T)^2 - E_{\Lambda}^2 }} \psi \end{array} \right],  \] 
$\vecc{\mu}$ is the block operator matrix given by 
\begin{equation} \vecc{\mu} =  \left[ \begin{array}{c} \vecc{0}^{T} \\ \vecc{\mu_1} \end{array} \right],  
\text{  with } \vecc{\mu}_1 = \mbox{diag}\left(\frac{a_n}{2 \Sigma_n}\right)_{n=0,1,\dots}, \end{equation}
$\vecc{\sigma}$ is the block operator  matrix given by 
\begin{equation} \vecc{\sigma} =  \left[ \begin{array}{c} \vecc{0}^{T} \\ \vecc{\Sigma} \end{array} \right],  \text{  with } \vecc{\Sigma} = \mbox{diag}\left(\Sigma_n \right)_{n=0,1,\dots} ,\end{equation} and 
\begin{equation} \vecc{W}^{\pi}_t = \left[  W^{\pi}_0(t)  \ \ W_1^{\pi}(t)  \ \cdots \right]^{T} \ \ \text{  and  }
\vecc{W}_{t}^{-\pi/2} = [ W_0^{-\pi/2}(t)  \ \ \  W^{-\pi/2}_{1}(t)  \  \ \cdots ]^{T}. \end{equation} 
In the above, the scalar-looking entries are really scalar multiples of appropriate identity operators.  In particular, $0$ denotes the zero operator on appropriate space.

%--------------------------------------------
\section{Formal Derivation of Limiting Equation} \label{derive}

We are interested in the limit as $\epsilon \to 0$ of \eqref{x_eqn}-\eqref{v_eqn}. These equations are similar to the ones studied in \cite{hottovy2015smoluchowski} and we adapt the techniques employed there and use the main results from quantum It\^o calculus outlined in Section \ref{background_qsde} to study the limit problem. 

In the limit $\epsilon \to 0$,  we expect that $\vecc{X}_{t}$ is a slow variable compared to $\vecc{V}_t$. In the following, we formally derive the limiting equation for the first component of $\vecc{X}_t$, i.e. the particle's position $X(t)$, in the limit $\epsilon \to 0$. To give meanings to our derivations, one considers the action of an operator, say $Z(t)$, on a vector of the form $\psi \otimes e(u)$, i.e. $Z(t) (\psi\otimes e(u))$,  where $ \psi \in \mathcal{H}_{S}$ and $e(u)$ is the exponential vector associated with $u \in L^2(\RR^+)\otimes \mathcal{K}$. We suppress this interpretation of operators in the following and work directly with the operators.

A rewriting of $\eqref{v_eqn}$ leads to:
\begin{equation} \vecc{V}_{t} dt = - \epsilon \vecc{\hat{\gamma}}^{-1}(X(t)) d\vecc{V}_{t} + \vecc{\hat{\gamma}}^{-1}(X(t)) \vecc{F}(X(t)) dt + \vecc{\hat{\gamma}}^{-1}(X(t)) \vecc{\sigma} d\vecc{W}^{-\pi/2}_{t}, \end{equation}
where $\vecc{\hat{\gamma}}^{-1}$ satisfies $\vecc{\hat{\gamma}}^{-1} \vecc{\hat{\gamma}} = \vecc{\hat{\gamma}} \vecc{\hat{\gamma}}^{-1} = I$ and can be verified to be given by the following block operator matrix:
\begin{equation} \vecc{\hat{\gamma}}^{-1}(X(t)) =  \left[ \begin{array}{cc}
\vecc{A}^{-1}(X(t)) & -\vecc{A}^{-1}(X(t))\vecc{B}(X(t)) \vecc{D}^{-1} \\ 
\vecc{0} & \vecc{D}^{-1} \end{array} \right],  \end{equation}
where \begin{equation} \vecc{A}^{-1}(X(t)) =  \left[ \begin{array}{cc} m_0 [f'(X(t))]^{-2} & -[a_0 f'(X(t))]^{-1}\\ m_0 [f'(X(t))]^{-1} & 0 
 \end{array} \right],  \end{equation}
\begin{equation} -\vecc{A}^{-1}(X(t))\vecc{B}(X(t)) \vecc{D}^{-1} =  \left[ \begin{array}{ccc} [a_0 f'(X(t))]^{-1} & [a_1 f'(X(t))]^{-1} & \cdots \\ a_0^{-1} & a_1^{-1} & \dots 
 \end{array} \right],  \end{equation} and \begin{equation} \vecc{D}^{-1}(X(t)) = \mbox{diag}(a_n^{-1})_{n=0,1,\dots}.\end{equation}
%\[ \vecc{\hat{\gamma}}^{-1}(X(t)) =  \left[ \begin{array}{cccccc} m_{0}[f'(X(t))^2]^{-1} & -\frac{1}{E_{\Lambda}} [f'(X(t))]^{-1} & \frac{1}{E_{\Lambda}} [f'(X(t))]^{-1} & \frac{1}{2\pi k_{B} T}[f'(X(t))]^{-1} & \dots & \frac{1}{2e N k_{B}T}[f'(X(t))]^{-1}  \\ m_{0}[f'(X(t))]^{-1} & 0 & \frac{1}{E_{\Lambda}} & \frac{1}{2\pi k_{B} T} & \dots & \frac{1}{2e N k_{B}T} \\ 0 & 0 & \frac{1}{E_{\Lambda}} & 0 & \dots & 0 \\  0 & 0 & 0  & \frac{1}{2 \pi k_{B} T} & 0 & 0 \\  0 & 0 & 0 & 0 & \ddots & 0 \\ 0 & 0 & 0 & 0 & 0 & \frac{1}{2 \pi n k_{B}T} \end{array} \right],  \] where we have assumed that  $[f'(X(t))]^{-1}f'(X(t)) = f'(X(t)) [f'(X(t))]^{-1} = I, \ [f'(X(t))^2]^{-1}f'(X(t)) = f'(X(t)) [f'(X(t))^2]^{-1} =  [f'(X(t))]^{-1}.$

As $d \vecc{X}_{t} = \vecc{V}_{t} dt + \epsilon \vecc{\mu} d\vecc{W}^{\pi}_{t}$, it follows that we can write $\vecc{X}_t$ in the integral form:
\begin{align}\vecc{X}_{t} &= \vecc{X} - \int_{0}^{t}   \epsilon \vecc{\hat{\gamma}}^{-1}(X(s)) d \vecc{V}_{s}   +  \int_{0}^{t}   \vecc{\hat{\gamma}}^{-1}(X(s)) \vecc{F}(X(s)) ds + \int_{0}^{t}  \vecc{\hat{\gamma}}^{-1}(X(s)) \vecc{\sigma} d\vecc{W}^{-\pi/2}_{s} \nonumber \\ 
&\ \ \ \ +   \epsilon \vecc{\mu}  (\vecc{W}^{\pi}_{t}-\vecc{W}^{\pi}). \end{align}

The only terms that depend explicitly on $\epsilon$ on the right hand side above are the second  term and the last term.  
Therefore, we study the asymptotic behavior of these terms as $\epsilon \to 0$. The last term will tend to zero as $\epsilon \to 0$. For the second term, we consider the components of the operator process $ \int_{0}^{t}   \epsilon \vecc{\hat{\gamma}}^{-1}(X(s)) d \vecc{V}_{s} = [D_1(t) \ D_2(t) \ \cdots ]^{T}$, where
\begin{align}
D_{1}(t) &= \int_{0}^{t} [f'(X(s))]^{-2} m_{0} \epsilon dV(s) - \int_{0}^{t} [a_0 f'(X(s))]^{-1} \epsilon dY(s) \nonumber \\  
&\ \ \ \  + \sum_{n=0}^{\infty} \int_{0}^{t} [a_n f'(X(s))]^{-1} \epsilon d\eta_n(s), \label{dt1} \\ 
D_{2}(t) &= \int_{0}^{t} [f'(X(s))]^{-1} m_{0} \epsilon dV(s) + \sum_{n=0}^{\infty} \int_{0}^{t} \frac{\epsilon}{a_n} d\eta_{n}(s), \\
D_{n+3}(t) &= \frac{\epsilon}{a_n} (\eta_{n}(t) - \eta_n), \ \ n = 0,1,2, \dots.
\end{align}
%Since \begin{equation} [f'(X(t))^2]^{-1} m_{0} \epsilon V(t)  \psi e(u)  - [f'(X_{0})^2]^{-1} m_{0} \epsilon V_{0} \psi e(u)  =   \int_{0}^{t} \frac{d}{ds}\left([f'(X(s))^2]^{-1}\right) m_{0} \epsilon V(s) ds \psi e(u)  + \int_{0}^{t} [f'(X(s))^2]^{-1} m_{0} \epsilon dV(s) \psi e(u) ,\begin{equation} where we have used the fundamental theorem of calculus and product rule,

In particular, the first component of $\vecc{X}_t$ is given by:
\begin{align} 
X(t) &= X - D_1(t) - \int_0^t [f'(X(s))]^{-2} U'(X(s)) ds + \frac{i}{2m_0} \int_0^t [f'(X(s))]^{-1} f''(X(s)) ds  \nonumber \\ 
&\ \ \ \ + \int_0^t \sum_{n=0}^{\infty} \frac{\Sigma_n}{a_n}  [f'(X(s))]^{-1} dW_{n}^{-\pi/2}(s). \label{ori_x}
\end{align}

Integrating by parts, we can write the first integral in \eqref{dt1} as: 
\begin{align}\int_{0}^{t} [f'(X(s))]^{-2} m_{0} \epsilon dV(s)  &= [f'(X(t))]^{-2} m_{0} \epsilon V(t) -[f'(X)]^{-2} m_{0} \epsilon V \nonumber \\ 
&\ \ \ \ \ - \int_{0}^{t} \frac{d}{ds}\left([f'(X(s))]^{-2}\right) m_{0} \epsilon V(s) ds.\end{align}

%Now as $[f'(X)^2]^{-1}$ commutes with $U(X)$, $f(X)$ and the bath variables, we have \begin{align*} \frac{d}{ds}\left([f'(X(s))^2]^{-1}\right) &= \frac{i}{\hbar}[H^{\epsilon}, [f'(X(s))^2]^{-1}] = \frac{i}{\hbar}\left[\frac{P^2}{2m_{0} \epsilon}, [f'(X(s))^2]^{-1} \right] = \frac{1}{2} \left\{ V(s), ([f'(X(s))^2]^{-1})' \right\} \\  &= ([f'(X(s))^2]^{-1})' V(s) - \frac{i}{4 m_{0}} ([f'(X(s))^2]^{-1})''.  \end{align*}  Note that in deriving the above, the scaling $\hbar/m_{0} = 1/m_{0}$ is again crucial. 

%We have  \begin{align*} &\int_{0}^{t} [f'(X(s))^2]^{-1} m_{0} \epsilon dV(s) \\  &= [f'(X(t))^2]^{-1} m_{0} \epsilon V(t) \\ &\ \ \ \ - [f'(X_{0})^2]^{-1} m_{0}  \epsilon V_{0} - \int_{0}^{t} \left( ([f'(X(s))^2]^{-1})' m_{0} \epsilon V(s)^2 - \frac{i}{4} ([f'(X(s))^2]^{-1})''  \epsilon V(s) \right) ds. \end{align*}

%Using \begin{align*} \frac{d}{ds}\left([f'(X(s))^2]^{-1}\right) \psi e(u) &= \frac{d}{ds}\left([f'(x_{s})^2]^{-1}\right) \psi(x) e(u) = \frac{d}{dx_{s}}\left([f'(x_{s})^2]^{-1}\right) v_{s} \psi(x) e(u) =: \frac{\partial}{\partial X(s)} ([f'(X(s))^2]^{-1}) V(s) \psi e(u),  \end{align*} 
Next, we make a remark on taking derivatives of operator-valued functions. Let $h(X(s))$ be a  function, depending on the position process $X(s)$, which can be expanded in a power series.  The formula for the derivative of the operator inverse reads
\begin{equation}\frac{d}{ds}([h(X(s))]^{-1}) = -[h(X(s))]^{-1} \left( \frac{d}{ds}[h(X(s))]  \right) [h(X(s))]^{-1}.\end{equation}
For $h(X(s)) = X(s)^p$, where $p=2,3,\dots$, rearranging the order to move $V(s)$ to the right, one obtains:
\begin{equation}\frac{d}{ds}X(s)^p = pX(s)^{p-1} V(s) - \frac{i \hbar}{m} X(s)^{p-2} c(p),\end{equation} where $c(p)$ is a constant depending on $p$. From this, one deduces: 
\begin{equation}\frac{d}{ds}[h(X(s))] = \left(\frac{\partial}{\partial X(s)} h(X(s)) \right) V(s) - \frac{i\hbar}{m} g(X(s)),\end{equation} for some function $g$, where $\frac{\partial}{\partial X(s)}$ denotes formal derivative  with respect to $X(s)$. Using this, it can be shown that, for some function $k$,
\begin{align}
\frac{d}{ds}([h(X(s))]^{-1}) &= -[h(X(s))]^{-1} \left( \frac{\partial}{\partial X(s)} h(X(s)) \right) [h(X(s))]^{-1} V(s) + \frac{i \hbar}{m} k(X(s)) \nonumber \\ 
&= \frac{\partial}{\partial X(s)} ([h(X(s))]^{-1}) V(s) +  \frac{i \hbar}{m} k(X(s)).
\end{align}
Note that $\hbar/m =1/m_0$ is independent of $\epsilon$ and the above remark allows us to apply the following chain rule for operators:
\begin{equation}\frac{d}{ds}\left([f'(X(s))]^{-2}\right) = \frac{\partial}{\partial X(s)} ([f'(X(s))]^{-2}) V(s) +  \frac{i}{m_0} l(X(s)),\end{equation}
for some function $l$, so that
 \begin{align} &\int_{0}^{t} [f'(X(s))]^{-2} m_{0} \epsilon dV(s) = [f'(X(t))]^{-2} m_{0} \epsilon V(t) - [f'(X)]^{-2} m_{0}  \epsilon V \nonumber \\ 
&\ \hspace{4cm} \ \ - \int_{0}^{t} \left( \frac{\partial}{\partial X(s)}[f'(X(s))]^{-2} \right) m_{0} \epsilon V(s)^2  ds -i \int_{0}^{t} l(X(s)) \epsilon V(s)  ds 
 \end{align} 
 
Guided by the estimates in the classical case \cite{LimWehr_Homog_NonMarkovian}, we expect that the terms in the above expression, which contain the momentum process, $\epsilon V(s)$, $s \in [0,t]$, tend to zero as $\epsilon \to 0$ and the terms containing the ``kinetic energy", $\epsilon V(s)^2$, are $O(1)$ as $\epsilon \to 0$. Physically, these statements can be justified by arguing that the momentum process is a fast variable that equilibrates rapidly and the equipartition theorem becomes valid in the considered limit, respectively. It is these contributions from $\epsilon V(s)^2$ that invalidate the naive procedure to obtain the limiting equation by simply setting $\epsilon$ to zero in the pre-limit equations; one  expects to obtain correction drift terms in the limiting equation for particle's position.

Similarly, we can repeat the above calculations and arguments for the other integral terms in \eqref{dt1}. We are thus left with the problem of deriving the limiting expressions for $\epsilon V(s)^2$, $\epsilon V(s) Y(s)$, $\epsilon V(s) \eta_{n}(s)$, $n=0, 1, \dots$ as $\epsilon \to 0$. 

To derive them, we apply quantum It\^o formula to \begin{equation} \epsilon^2 \vecc{V}_{s} \vecc{V}_{s}^{T}    =  \epsilon^2 \left[ \begin{array}{ccccc} V(s)^2 & V(s)Y(s) & \dots &  V(s)  \eta_{N}(s) & \cdots \\ Y(s)  V(s)  & Y(s)^2 & \dots & Y(s)  \eta_{N}(s) & \cdots \\   \vdots & \vdots & \ddots & \vdots  \\   \eta_{N}(s) V(s) &   \eta_{N}(s) Y(s) & \dots &  \eta^2_{N} (s) & \dots \\ \vdots & \vdots &  & \vdots & \ddots \end{array} \right],  \end{equation}
where the entries should be interpreted as tensor products of operators.

This gives:
\begin{align}
d[(\epsilon \vecc{V}_{s})(\epsilon \vecc{V}_{s}^{T})] &= d[\epsilon \vecc{V}_{s}] \epsilon \vecc{V}_{s}^{T} + \epsilon \vecc{V}_{s} d[\epsilon \vecc{V}_{s}^{T}] + d[\epsilon \vecc{V}_{s}] d[\epsilon \vecc{V}_{s}^{T}] \nonumber \\ 
&= [-\vecc{\hat{\gamma}}(X(s)) \vecc{V}_{s} + \vecc{F}(X(s)) + \vecc{\sigma} d\vecc{W}^{-\pi/2}_{s}] \epsilon \vecc{V}_{s}^{T} ds \nonumber \\
&\ \ \ \ + \epsilon \vecc{V}_{s} [-\vecc{\hat{\gamma}}(X(s)) \vecc{V}_{s} + \vecc{F}(X(s)) + \vecc{\sigma}d\vecc{W}^{-\pi/2}_{s}]^{T} ds + \vecc{\sigma} \vecc{\sigma}^{T} ds, 
\end{align}
where we have used the quantum It\^o formula \eqref{qif} to compute: \begin{equation}d\vecc{W}^{-\pi/2}_{s}  d(\vecc{W}^{-\pi/2}_{s})^{T} = \vecc{I} ds.\end{equation}
Rearranging, we obtain the following {\it operator Lyapunov equation} \cite{bhatia1997and,rosenblum1956operator}: \begin{equation}\vecc{\hat{T}}(\vecc{J}) := \vecc{\Gamma} \vecc{J} + \vecc{J} \vecc{\Gamma}^{T} = \vecc{B}_1+\vecc{B}_2+\vecc{B}_3, \end{equation} 
where \begin{equation}\vecc{\Gamma} = \vecc{\hat{\gamma}}(X(s)), \ \ \  \vecc{J} = \epsilon \vecc{V}_{s}  \vecc{V}_{s}^{T} ds, \ \ \ \vecc{B}_1 = - d[\epsilon^2 \vecc{V}_{s} \vecc{V}_{s}^{T}],\end{equation} 
\begin{equation}\vecc{B}_2 =  \epsilon[(\vecc{\sigma} d\vecc{W}^{-\pi/2}_{s} + \vecc{F}(X(s))) \vecc{V}_{s}^{T} + \vecc{V}_{s} (d(\vecc{W}^{-\pi/2}_{s})^{T} \vecc{\sigma}^{T} + \vecc{F}(X(s))^{T}) ] ds, \ \ \ \ \vecc{B}_3 =  \vecc{\sigma} \vecc{\sigma}^{T} ds.\end{equation}
The formal solution to this equation can be written as
\begin{equation}\vecc{J} = \vecc{\hat{T}}^{-1}(\vecc{B}_1) + \vecc{\hat{T}}^{-1}(\vecc{B}_2) + \vecc{\hat{T}}^{-1}(\vecc{B}_3).\end{equation}
%\cite{bhatia1997and,rosenblum1956operator}:\begin{equation}\vecc{J} =   \int_{0}^{\infty} e^{-\vecc{\Gamma} y} \vecc{B}_1 e^{-\vecc{\Gamma}^{T}y} dy  +\int_{0}^{\infty} e^{-\vecc{\Gamma} y} \vecc{B}_2 e^{-\vecc{\Gamma}^{T}y} dy  +\int_{0}^{\infty} e^{-\vecc{\Gamma} y} \vecc{B}_3 e^{-\vecc{\Gamma}^{T}y} dy.\begin{equation} 
By previous arguments on the asymptotic behavior of the momentum process, we expect that $\vecc{\hat{T}}^{-1}(\vecc{B}_1)$ and $\vecc{\hat{T}}^{-1}(\vecc{B}_2)$ tend to zero as $\epsilon \to 0$. 

Therefore, in the limit $\epsilon \to 0$, $\epsilon \vecc{V}_{s} \vecc{V}_{s}^{T}$ converges to the solution, $\vecc{\bar{J}}$, of the operator  Lyapunov equation:
\begin{equation}\vecc{\hat{\gamma}}(\bar{X}(s)) \vecc{\bar{J}} + \vecc{\bar{J}} \vecc{\hat{\gamma}}(\bar{X}(s))^{T}  = \vecc{\sigma} \vecc{\sigma}^{T}.\end{equation}
Solving the above equation for $\vecc{\bar{J}}$ allows us to extract the limits of $\epsilon V(s)^2$, $\epsilon V(s) Y(s)$ and $\epsilon V(s) \eta_n(s)$, $n=0,1,\dots$, which we denote by $J_{1,1}$, $J_{1,2}$, $J_{1,n+3}$ for $n=0,1,\dots$, respectively.  We refer to Appendix \ref{app_lyapunov} for the solution of this equation.
This is what we need to determine the asymptotic behavior of $X(t)$ in \eqref{ori_x} as $\epsilon \to 0$.  
The following limiting equation for $\bar{X}(t)$ is the main result of this paper. It is derived for a large class of non-Markovian QBM with inhomogeneous damping and diffusion and is valid for positive temperature. \\ 

{\bf The Main Result.} In the limit $\epsilon \to 0$, the particle's position, $X(t)$, converges to the solution, $\bar{X}(t)$, of the following equation:  
\begin{align} \label{mainresult} d\bar{X}(t) &= - [f'(\bar{X}(t))]^{-2} U'(\bar{X}(t)) dt + \frac{i}{2m_{0}} [f'(\bar{X}(t))]^{-1} f''(\bar{X}(t)) dt + S(\bar{X}(t)) dt  \nonumber  \\  &\ \ \ + [f'(\bar{X}(t))]^{-1} \left( \sqrt{a_{0}\cot\left(\frac{a_{0}}{2k_B T} \right)} dW_0^{-\pi/2}(t) +  \sum_{n=1}^{\infty} \sqrt{\frac{4a_{0}^2 k_B T}{a_n^2 - a_{0}^2}} dW_n^{-\pi/2}(t) \right) ,\end{align}  where $S(\bar{X}(t))$ is the {\it quantum noise-induced drift}, arising in the limit of simultaneously vanishing inertial, memory, noise correlation and quantum time scales, given by: \begin{align} \label{qnids} S(\bar{X}) &= \left(\frac{\partial}{\partial \bar{X}}[f']^{-2} \right) \frac{a_{0}}{2}\cot\left( \frac{a_{0}}{2k_{B}T} \right) \nonumber \\  &\ \ \  + \left(\frac{\partial}{\partial \bar{X}}[f']^{-2} \right) \sum_{n=1}^{\infty} \left\{ \frac{2 k_B T a_{0}^2 }{a_n^2 - a_{0}^2} \left(I + \frac{a_{n}}{m_0 a_{0}(a_{n}+a_{0})} (f')^2 \right) \left(I+ \frac{a_{0}}{m_{0}a_{n}(a_{n}+a_{0})}(f')^2  \right)^{-1} \right\} \nonumber \\  &\ \ \ - \left(\frac{\partial}{\partial \bar{X}} [f']^{-1} \right) \sum_{n=1}^{\infty} \left\{ \frac{2k_{B}T a_{0}}{m_0 a_n (a_n+a_{0})} \left(I+ \frac{a_{0}}{m_0 a_{n}(a_{n}+a_{0})} (f')^2 \right)^{-1} \right\} f',   \end{align} where  $a_0 =E_{\Lambda}$, $a_{n} = 2 \pi n k_{B} T$ and  the $W^{-\pi/2}_{n}(t) = i(A_n(t) - A^\dagger_n(t))$, $n=0,1,\dots$, are independent quantum Wiener processes introduced in \eqref{qwp} of Section \ref{background_qsde}.

We make a few remarks about the limiting equation \eqref{mainresult}. 

First, as in the classical case, it contains drift correction terms induced by  vanishing of all the characteristic time scales. The presence of such noise-induced drift is a consequence of nonlinear coupling in the QBM model. Expanding $S(\bar{X})$ about large $T$, we see that:
\begin{equation}S(\bar{X}) =  \frac{\partial}{\partial \bar{X}}\left([f'(\bar{X})]^{-2} \right) k_{B}T + O(1/T),\end{equation} and so the form of the zeroth order contribution coincides with the classical noise-induced drift obtained in the case of equilibrium bath where the Einstein's relation is satisfied (c.f. (101) in \cite{hottovy2015smoluchowski} with $D=[f']^{-2} k_B T$). Moreover, the zeroth order contribution after expanding in large $T$ for the noise terms reads \begin{equation}[f']^{-1}\sqrt{2k_BT} dW_0^{-\pi/2}(t),\end{equation}
whose form (modulo the quantum nature of the noise $W_0^{-\pi/2}$) is identical to that in the classical result. Therefore, we see that our quantum noise-induced drift consists of a classical counterpart (the first term in the formula for $S(\bar{X})$ above), which reduces to the classical noise-induced drift in the high temperature regime, and also several drift terms that are purely quantum in origin. The latter drifts depend  explicitly on both bath parameters $a_{n}$ and $E_{\Lambda}$. Note that, in contrast to the classical results and to the linear coupling case, we have an additional term (the contribution involving the imaginary number) in the limiting equation due to inhomogeneous nature of the bath.

Second, in the linear coupling ($f(\bar{X}) = \bar{X})$ case, the limiting equation reduces to:
\begin{equation}d\bar{X}(t) = -  U'(\bar{X}(t)) dt +  \sqrt{E_{\Lambda}\cot\left(\frac{E_{\Lambda}}{2k_B T} \right)} dW_0^{-\pi/2}(t) +  \sum_{n=1}^{\infty} \sqrt{\frac{4E_{\Lambda}^2 k_B T}{a_n^2 - E_{\Lambda}^2}} dW_n^{-\pi/2}(t). \end{equation}
In contrast to the results obtained in the literature (see for instance, eqn. (14) in \cite{Ankerhold1}), the limiting equation is not a classical SDE, but a QSDE driven by quantum thermal noises. At low temperature, all the quantum noise terms in the above expression contribute significantly to the limiting dynamics.  While this is in agreement with the finding  in \cite{Ankerhold1} that quantum fluctuations play an important role at low temperatures, the detailed expression of this role obtained here is different.  On the other note, in this special case $S(\bar{X})=0$, so drift correction terms are absent in the limiting equation and the formal procedure of setting $\epsilon$ to zero in \eqref{res_hl1}-\eqref{res_hl6} yields a correct limiting equation.

Third, to demonstrate the relations between the noise coefficients and the parameters in  the noise-induced drifts, one can rewrite:
\begin{align}
d\bar{X}(t) &= - [f'(\bar{X}(t))]^{-2} U'(\bar{X}(t)) dt + \frac{i}{2m_{0}} [f'(\bar{X}(t))]^{-1} f''(\bar{X}(t)) dt + S(\bar{X}(t)) dt   \nonumber \\ 
&\ \ \ + \sqrt{4 k_{B}T} [f'(\bar{X}(t))]^{-1}  \sum_{n=0}^{\infty} \beta_n  dW_n^{-\pi/2}(t)  ,
 \end{align} 
where 
\begin{align}
S(\bar{X}) &= 2k_B T \bigg[ \frac{\partial}{\partial \bar{X}}\left([f']^{-2} \right) \beta_0^2  + \frac{\partial}{\partial \bar{X}}\left([f']^{-2} \right) \sum_{n=1}^{\infty} \left\{ \beta_n^2 \left(I +  \frac{\beta_{n}^2}{m_0 c_n} (f')^2 \right) \left(I+  \frac{\beta_n^2}{m_{0}d_n}  (f')^2  \right)^{-1} \right\} \nonumber \\ 
&\ \ \ - \frac{\partial}{\partial \bar{X}} \left([f']^{-1} \right) \sum_{n=1}^{\infty} \left\{ \frac{ \beta_n^2}{m_0 d_n} \left(I+ \frac{\beta_n^2}{m_0 d_n} (f')^2 \right)^{-1} \right\} f'\bigg], 
\end{align}
with   
\begin{equation}\beta_0 = \sqrt{\frac{a_0}{4k_B T} \cot\left(\frac{a_0}{2k_B T} \right)}, \ \ \beta_n = \sqrt{\frac{a_0^2}{a_n^2-a_0^2}}, \ \  d_n = a_n \frac{a_0}{a_n-a_0},  \ \ c_n = \frac{a_0^2}{a_n} \frac{a_0}{a_n-a_0}.   \end{equation} 
The formulae for the noise parameters $\beta_n$ ($n =1,2,\dots$)  resembles those derived in \cite{attal2007langevin}. In particular, $\beta_n^2$ can be written as $1/(e^{\epsilon(n)/k_B T}-1)$, with $\epsilon(n) = 2k_B T \ln(a_n/a_0) > 0$ and similarly for $d_{n}/a_n = a_n c_n/a_0^2$. Therefore, information about  expected number of bosons in an energy state of energy $\epsilon(n)$ is encoded in the noise coefficients of the limiting equation and, more importantly, since the quantum noise-induced drifts depend explicitly on $\beta_n^2$, they also encode such information about the heat bath.

Fourth, for the nonlinear coupling case the limiting equation does not describe a Markovian dynamics and so cannot be cast as a QSDE in a H-P form, due to the presence of nonzero contribution containing the imaginary expression in the equation. On the other hand, for the linear coupling case, the limiting equation can be cast into a H-P QSDE. However, the H-P form is not unique since the associated effective Hamiltonian and Lindblad operators can only be deduced from the form of Heisenberg-Langevin equation for a single observable and the Lindblad form is invariant under certain transformations of the Hamiltonian and Lindblad operators.  %This illustrates the shortcomings of our method of studying the effective dynamics via the Heisenberg-Langevin approach. 
One way to choose a Lindblad form is to argue as follows. In the usual weak coupling limit the effective dynamics would be a Markovian non-dissipative dynamics with the Lindblad operator given by $L = X$ and the effective system Hamiltonian $H_{eff}$ equal $H_{S}$ (plus possibly a  correction term proportional to $X^2$). Since here we are taking small mass limit together with a white noise limit, one would expect to obtain Markovian dynamics associated with a modified $L$ and a modified $H_{e}$. In this way, one {\it postulates}  the Lindblad operators to be:  
\begin{align}   L_{n} = \bar{X} - \sqrt{4k_B T} \beta_n P/\hbar, \ \ \ n=0,1,2,\dots \end{align}  and the effective system Hamiltonian to be: \begin{align}  H_{e} &= \{ - U'(\bar{X}), P\}/2. \end{align}

%\begin{align}   L_{n} = f(\bar{X}) - \sqrt{4k_B T} \beta_n [f'(\bar{X})]^{-1} P/\hbar, \ \ \ n=0,1,2,\dots \end{align}  and the effective system Hamiltonian to be: \begin{align}  H_{e} &= \left( -[f'(\bar{X})^2]^{-1} U'(\bar{X}) + \frac{i}{2m_{0}}[f'(\bar{X})]^{-1} f''(\bar{X}) + S(\bar{X}) + \sum_{n=0}^{\infty} 4k_B T \beta_n^2 [f'(\bar{X})]^{-3}f''(\bar{X}) \right)P. \end{align} 

%With the above postulated Lindblad operators and Hamiltonian, one can then obtain a stochastic Schrodinger equation for the evolution operator and a master equation for the reduced density operator describing the effective dynamics. One can also study the mean values of observables and their stationary solutions \cite{lampo2016lindblad}. 

%-------------------------------------------

\section{Conclusions and Final Remarks} \label{final}
In this paper, we  study the small mass limit of  QBM model using a quantum stochastic calculus approach, extending analogous studies for classical models. More precisely, in the limit considered, the particle's momentum is a fast variable that can be adiabatically eliminated to obtain a reduced dynamics described by evolution of position alone, while at the same time %the memory effects are coarse-grained away. 
%This is the limit where distribution for the position is infinitely squeezed.  
memory effects reduce to additional drifts.
Our main result is derivation of the limiting equation \eqref{mainresult} for the particle's position. This equation exhibits strong quantum effects.  It is driven by thermal noises which are linear combinations of H-P fundamental processes. Its most important feature is the presence of {\it quantum noise-induced drift} given in eqn. \eqref{qnids}, which corrects the equation one would obtain by naively setting $\epsilon$ to zero in the pre-limit equations. This correction  consists of terms which have classical counterparts, as well as drifts that are purely quantum in origin. 

We expect that such quantum noise-induced drifts will lead to interesting effects in experimental studies of small mass quantum system at low temperatures, such as an impurity in a ultracold Bose gas. 
In \cite{Cugliandolo2012,Bonart2013,Massignan2015,Lampo2017}, it has been shown that the Hamiltonian of this system may be cast in the form of a QBM model. Here, the impurity plays the role of the Brownian particle, while the environment is represented by the Bogoliubov excitations of the gas. 
In general, the coupling between the impurity and the bath shows a nonlinear dependence on the position of the former. 
Accordingly, this system is a good candidate to detect a quantum noise-induced drift. 
However, the spectral density of an impurity in a Bose gas cannot be reduced to an Ohmic one. 
For instance, in \cite{Lampo2017} it has been shown that, in the case in which the gas is homogeneous, i.e. its density is space-independent, the spectral density shows the following behavior:
\begin{equation}
J(\omega)\sim\omega^{d+2},
\end{equation} 
where $d$ is the dimension of the system. Therefore,
it would be interesting to extend the present study to a QBM model where the bath spectral density is different from the one considered here, in particular the non-Ohmic ones \cite{ford2006anomalous,efimkin2016non}. 
We will leave these further explorations to future work.

\section*{Acknowledgements}
S. Lim and J. Wehr were partially supported by NSF grant DMS 1615045. This work has been funded by a scholarship from the Programa M\'{a}sters d'Excel-l\'{e}ncia of the Fundaci\'{o} Catalunya-La Pedrera, ERC Advanced Grant OSYRIS, EU IP SIQS, EU PRO QUIC, EU STREP EQuaM (FP7/2007-2013, No. 323714), Fundaci\'o Cellex, the Spanish MINECO (SEVERO OCHOA GRANT SEV-2015-0522,  FOQUS FIS2013-46768, FISICATEAMO FIS2016-79508-P), and the Generalitat de Catalunya (SGR 874 and CERCA/Program).

\bibliographystyle{apsrev4-1_our_style}
\bibliography{ref}

%---------------------------------------------
\section*{Appendices}
\appendix

\section{Derivation of Heisenberg Equations for Particle's Observables} \label{appendix_qle}

In this appendix we derive  equations \eqref{qle1}-\eqref{qle2}. Let
\begin{equation}b(\omega) = \sqrt{\frac{\omega}{2\hbar}}\left(x(\omega)+\frac{i}{\omega}p(\omega) \right), \ \ \ \  b^{\dagger}(\omega) = \sqrt{\frac{\omega}{2\hbar}}\left(x(\omega)-\frac{i}{\omega}p(\omega) \right),  \end{equation}
\begin{equation}
[x(\omega),p(\omega')] = i\hbar \delta(\omega-\omega')I,\end{equation} where we have normalized the masses of all bath oscillators. 

The Heisenberg equation of motion gives \begin{align} \dot{X}(t) &= \frac{i}{\hbar} [H, X(t)] = \frac{P(t)}{m}, \\ 
\dot{P}(t) &= \frac{i}{\hbar} [H, P(t)] \nonumber \\ 
&= -U'(X(t)) + f'(X(t)) \int_{\RR^{+}} d\omega c(\omega) \sqrt{\frac{2\omega}{\hbar}} x_{t}(\omega) - 2f(X(t)) f'(X(t)) \int_{\RR^{+}} r(\omega) d\omega, \\  \dot{x}_{t}(\omega) &= \frac{i}{\hbar} [H, x_{t}(\omega)] = p_{t}(\omega), \ \ \omega \in \RR^{+}, \\  \dot{p}_{t}(\omega) &= \frac{i}{\hbar} [H, p_{t}(\omega)] = - \omega^2 x_{t}(\omega) + \sqrt{\frac{2\omega}{\hbar}} c(\omega) f(X(t)), \ \ \omega \in \RR^{+},  \end{align} where $r(\omega) = |c(\omega)|^2/(\hbar \omega)$ and $f'(X)= [f(X),P ]/(i\hbar)$. 

Next we eliminate the bath degrees of freedom from the equations for $X(t)$ and $P(t)$. Solving for $x_{t}(\omega)$, $\omega \in \RR^{+}$, gives: \begin{equation}x_{t}(\omega) = \underbrace{x_{0}(\omega) \cos(\omega t) + p_{0}(\omega) \frac{\sin(\omega  t)}{\omega}}_{x^{0}_{t}(\omega)} + \int_{0}^{t} \frac{\sin( \omega (t-s))}{\omega} \sqrt{\frac{2\omega}{\hbar}} c(\omega) f(X(s)) ds. \end{equation} Substituting this into the equation for $P(t)$ results in:  \begin{align} \dot{P}(t) &= -U'(X(t)) + f'(X(t)) \int_{\RR^{+}} d\omega  c(\omega) \sqrt{\frac{2\omega}{\hbar}} x^{0}_{t}(\omega) \nonumber \\ 
&\ \ \  +  \frac{2}{\hbar} f'(X(t)) \int_{\RR^{+}} d\omega  |c(\omega)|^2  \int_{0}^{t} ds \sin(\omega  (t-s)) f(X(s)) - 2 f(X(t)) f'(X(t)) \int_{\RR^{+}} d\omega r(\omega). \end{align} 

Using integration by parts, we obtain \begin{equation}\int_{0}^{t} ds \sin( \omega(t-s)) f(X(s)) = \frac{f(X(t))}{\omega} - f(X)\frac{\cos(\omega t)}{\omega} - \int_{0}^{t} \frac{\cos(\omega(t-s))}{\omega} \frac{d}{ds}\left(f(X(s)) \right)ds   \end{equation} and therefore, \begin{align} \dot{P}(t) &= -U'(X(t)) + f'(X(t)) \underbrace{\int_{\RR^{+}} d\omega  c(\omega) (b^{\dagger}_t(\omega) + b_t(\omega) ) }_{\zeta(t)} \nonumber \\ 
&\ \ \ \ -  f'(X(t))  \int_{0}^{t} ds \underbrace{\int_{\RR^{+}} d\omega  2 r(\omega) \cos(\omega(t-s))}_{\kappa(t-s)}  \frac{d}{ds}\left(f(X(s)) \right) \nonumber \\ 
&\ \ \ \ -  f'(X(t)) f(X) \underbrace{\int_{\RR^{+}} d\omega  2r(\omega)  \cos(\omega t)}_{\kappa(t)},  \end{align} where \begin{equation}\frac{d}{ds}\left(f(X(s)) \right) = \frac{i}{\hbar}[H, f(X(s))] = \frac{\{ f'(X(s)), P(s)\}}{2 m},\end{equation} $b_t(\omega) = b(\omega)e^{-i\omega t}$, $b_t^{\dagger}(\omega) = b^{\dagger}(\omega) e^{i\omega t}$ and $\{\cdot, \cdot\}$ denotes anti-commutator.

%\rmk (On QLEs) The QLEs are operator equations that acts in the full Hilbert space of system and bath. The coupling between system and environment also implies an entanglement upon time evolution even for the case of an initially factorizing full density matrix. Together with the commutator property of quantum Brownian motion, the reduced, dissipative dynamics of the position operator and momentum operator should obey the Heisenberg uncertainty relation for all times: thus simply replacing the quantum noise by a classical noise would violate the uncertainty relation. Also, the fact that the QLE acts in full Hilbert space of system and environment needs to be distinguished from the classical case of a generalized Langevin equation. There, the stochastic dynamics acts solely on the state space of the system dynamics with the (classical) noise properties specified a priori.

\section{Solving the Operator Lyapunov Equation} \label{app_lyapunov}
We outline the derivation of the solution, $\vecc{\bar{J}}$, to the operator  Lyapunov equation:
\begin{equation}\vecc{\hat{\gamma}}(\bar{X}(s)) \vecc{\bar{J}} + \vecc{\bar{J}} \vecc{\hat{\gamma}}(\bar{X}(s))^{T}  = \vecc{\sigma} \vecc{\sigma}^{T},\end{equation}
where $\vecc{\hat{\gamma}}$ and $\vecc{\sigma} $ are block operator matrices, defined in Section \ref{rescaledmodel}. First, we observe that upon taking transpose on both sides of the equation, we have $\vecc{\hat{\gamma}}(\bar{X}(s)) \vecc{\bar{J}}^{T} + \vecc{\bar{J}}^{T} \vecc{\hat{\gamma}}(\bar{X}(s))^{T}  = \vecc{\sigma} \vecc{\sigma}^{T}$, so uniqueness of the solution implies $\vecc{\bar{J}} = \vecc{\bar{J}}^{T}$, i.e. $J_{k,l} = J_{l,k}$ for all $k,l$.

We write $\vecc{\bar{J}}$ in the block-structure form: 
\begin{equation} \vecc{\bar{J}} =  \left[ \begin{array}{cc} \vecc{J}_1 & \vecc{J}_2 \\ \vecc{J}_2^{T} & \vecc{J}_4 
 \end{array} \right],\end{equation} 
where  
\begin{equation} \vecc{J}_1 =  \left[ \begin{array}{cc} J_{1,1} & J_{1,2} \\ J_{1,2} & J_{2,2} 
 \end{array} \right], \ \ \ \vecc{J}_2 =  \left[ \begin{array}{ccc} J_{1,3} & J_{1,4} & \cdots \\ J_{2,3} & J_{2,4} & \cdots 
 \end{array} \right] \ 
\text{ and  } \ \vecc{J}_4 =  \left[ \begin{array}{ccc} J_{3,3} & J_{3,4} & \cdots \\ J_{4,3} & J_{4,4} &  \cdots \\
\vdots & \vdots & \ddots
\end{array} \right].\end{equation}

Working out the matrix  multiplications of the block operator  matrices in the equation gives
\begin{equation}\vecc{J}_4 = \frac{1}{2} \vecc{D}^{-1} \vecc{\Sigma}^2,\end{equation} which is a diagonal block operator matrix, 
and the following Sylvester-type equations:
\begin{align}
\vecc{A} \vecc{J}_2 + \vecc{J}_2 \vecc{D} &= - \frac{1}{2} \vecc{B} \vecc{D}^{-1} \vecc{\Sigma}^2, \label{s1} \\ 
\vecc{A} \vecc{J}_1 + \vecc{J}_1 \vecc{A}^{T} &= -\vecc{B} \vecc{J}_2^{T} - \vecc{J}_2 \vecc{B}^{T}. \label{s2}
\end{align}

Eqn. \eqref{s1} gives a system of linear equations for $J_{1,n+3}$ and $J_{2,n+3}$, for $n=0,1,\dots$:
\begin{align}
\frac{f'}{m_0} J_{2,n+3} + a_n J_{1,n+3} &= \frac{f'}{2m_0} \frac{\Sigma_n^2}{a_n}, \\ 
-a_0 f' J_{1,n+3} + (a_0 + a_n) J_{2,n+3} &= 0,
\end{align}
which has the solution:
\begin{align}
J_{2,n+3} &= \frac{\Sigma_n^2}{2m_0} \frac{a_0}{a_n^2(a_0+a_n)} \left[I+\frac{a_0}{m_0 a_n(a_0+a_n)} (f')^2  \right]^{-1} (f')^2, \label{s3} \\ 
J_{1,n+3} &= \frac{\Sigma_n^2}{2m_0 a_n^2}   \left[I+\frac{a_0}{m_0 a_n(a_0+a_n)} (f')^2  \right]^{-1} f', \label{s4}
\end{align}
where we have used the fact that $h(X)g(X) = g(X) h(X)$ for any functions $g$, $h$. 
Similarly, eqn. \eqref{s2} gives:
\begin{align}
J_{1,2} &= \sum_{n=0}^{\infty} J_{1,n+3}, \ \ \ J_{2,2} = \frac{1}{2} \{f', \sum_{n=0}^{\infty} J_{1,n+3} \}, \\ 
J_{1,1} &= \left((f')^{-1} + \frac{1}{m_0 a_0} f'\right)  \sum_{n=0}^{\infty} J_{1,n+3} - \frac{1}{m_0 a_0} \sum_{n=0}^{\infty} J_{2,n+3}.
\end{align}
Substituting the expressions for $J_{2,n+3}$ and $J_{1,n+3}$ from \eqref{s3}-\eqref{s4} into the above equation gives the formula for $J_{1,2}$, $J_{2,2}$ and $J_{1,1}$. In particular,
\begin{equation}J_{1,1} =  \sum_{n=0}^{\infty} \left\{ \frac{\Sigma_n^2}{2m_0 a_n^2} \left[I+\frac{a_n}{m_0 a_0(a_0+a_n)} (f')^2 \right] \left[I + \frac{a_0}{m_0 a_n(a_0+a_n)} (f')^2 \right]^{-1} \right\}.\end{equation}

We remark that upon taking the limit, the contributions involving the $J_{1,2}$ and $J_{1,3}$ cancel each other, as in the classical situation, and so the contributions coming from $J_{1,n}$ ($n \geq 4$) are indeed correction drift terms induced by purely quantum noises. 

\end{document}